\documentclass[11pt,a4paper]{article}

\usepackage[a4paper,margin=24mm,headheight=14pt]{geometry}
\usepackage{amsmath,amssymb,bm,mathtools}
\usepackage{graphicx}
\usepackage{booktabs,tabularx,array,multirow}
\usepackage{url}
\usepackage{rotating}
\usepackage{microtype}
\usepackage{xcolor}                         
\usepackage{caption}                         
\usepackage{float}                           
\usepackage{placeins}                        
\usepackage{enumitem}                        
\usepackage{siunitx}                         
\usepackage[numbers,sort&compress]{natbib}
\usepackage{hyperref}                       
\usepackage{footnotehyper}                  
\usepackage[nameinlink,noabbrev]{cleveref}

\graphicspath{{figures/large_field/}{figures/small_field/}}
\hypersetup{colorlinks=true,linkcolor=blue,citecolor=blue,urlcolor=blue}
\allowdisplaybreaks
\setlist[itemize]{leftmargin=1.5em,itemsep=0.2em,topsep=0.2em}
\setlist[enumerate]{leftmargin=1.7em,itemsep=0.25em,topsep=0.25em}
\newcommand{\mpl}{M_{\rm P}}
\newcommand{\CW}{\mathrm{CW}}

\newcommand{\eff}{\mathrm{eff}}
\newcommand{\dd}{\mathrm{d}}
\newcommand{\As}{A_{\rm s}}
\newcommand{\Pt}{\mathcal P_{\rm T}}
\newcommand{\Prm}{\mathcal P_{\mathcal R}}

\newcommand{\lambdat}{\widetilde{\lambda}_0}

\newcommand{\ns}{n_{\rm s}}

\newcommand{\cL}{\mathcal L}

\title{\textbf{Coleman-Weinberg potential, unimodular scalar-tensor gravity and cosmic inflation in Einstein frame}}
\author{Amirhossein Firouzi\hyperlink{fn:e1}{\textsuperscript{1}}\hypertarget{src:e1}{}
 \ and Hossein Ghaffarnejad\hyperlink{fn:e2}{\textsuperscript{2}}\hypertarget{src:e2}{}\\[0.4em]
 \small Faculty of Physics, Semnan University, Semnan, Iran, P.O. Box: 35131--19111}
\date{}

\begin{document}
\maketitle

\footnotetext[1]{\hypertarget{fn:e1}{}\hyperlink{src:e1}{E-mail address:}
  \href{mailto:hghafarnejad@semnan.ac.ir}{amirhosseinfirouzi@semnan.ac.ir}}
\footnotetext[2]{\hypertarget{fn:e2}{}\hyperlink{src:e2}{E-mail address:}
  \href{mailto:amirhossein_firouzi@semnan.ac.ir}{hghafarnejad@semnan.ac.ir}}
\begin{abstract}
We study Coleman--Weinberg inflation in unimodular gravity, retaining the integration constant associated with the unimodular constraint. The scalar field has a vacuum-shifted nonminimal curvature coupling in the original formulation which after transformation to the Einstein frame the gravitational sector becomes minimally coupled. We derive the Einstein-frame scalar kinetic factor and establish the constancy of the Lagrange multiplier using the scalar equation and the contracted Bianchi identity. We investigate both large- and small-field inflationary branches using the potential slow-roll approximation. Our work is optimized for $N=50$, $60$, and $70$, and the integration constant is varied with the scalar amplitude renormalized self-consistently at each point. We compare the resulting predictions with the weighted Planck, ACT-lite, BICEP/Keck, lensing, and BAO posterior samples used in this analysis and try to find out how the variation of unimodular integration constant can modify inflationary parameters.
\end{abstract}


\section{Introduction}

Inflation is a period of accelerated expansion in the early Universe, defined by $\ddot a>0$. Its expansion need not be exactly exponential: quasi-de Sitter evolution is common, while power-law expansion, such as $a(t)\propto t^p$ with $p>1$, also describes inflationary solutions. The inflationary paradigm was developed to account for the horizon, flatness, and magnetic-monopole problems of the hot Big Bang cosmology, and it provides a causal mechanism for producing the primordial perturbations from which the cosmic microwave background anisotropies and the large-scale structure subsequently form \citep{Starobinsky1980,Guth1981,Linde1982,AlbrechtSteinhardt1982,MukhanovChibisov1981}. The comparison between inflationary predictions and observations has evolved from qualitative model building into a precision test. In particular, the scalar amplitude, scalar spectral index, and tensor-to-scalar ratio constrain both the height and the shape of the inflationary potential \citep{LiddleParsonsBarrow1994,StewartLyth1993,LiddleLeach2003,Planck2020,BK2021}.

A useful way to broaden the possible inflationary dynamics is to modify the gravitational sector by coupling the inflaton to the Ricci scalar. Nonminimal scalar--curvature interactions arise naturally in quantum field theory in curved spacetime and play a central role in Higgs inflation and in radiatively corrected scalar models \citep{FakirUnruh1990,BezrukovShaposhnikov2008,Kaiser2010}. A second, conceptually distinct modification is to impose the unimodular condition on the spacetime volume element. In unimodular gravity (UG), the cosmological constant can arise as an integration constant of the gravitational equations rather than as a parameter inserted in the action. The purpose of the present work is to combine these ingredients in a Coleman--Weinberg model and, rather than setting the unimodular integration constant to zero, determine whether its contribution can improve the inflationary predictions.

The historical origin of UG is usually traced to Einstein's 1919 discussion of a restriction on the determinant of the metric, followed by Pauli's later discussion of the subject and the subsequent field-theoretic developments of the early 1970s and 1980s \citep{Einstein1919,PauliRelativity,AndersonFinkelstein1971,VanDerBij1982,BuchmullerDragon1988,BuchmullerDragon1989}. In its modern form, UG is a gravitational theory in which the spacetime volume element is fixed. In a coordinate system adapted to the prescribed volume form, the condition is commonly written as
\begin{equation}
\sqrt{-g}=\varepsilon_0(x),
\label{eq:unimodular_constraint_general}
\end{equation}
where $\varepsilon_0(x)$ is a fixed positive density. In a convenient unimodular coordinate chart we set $\varepsilon_0=1$. The restriction is therefore a statement about the metric volume element relative to a chosen volume form, not a condition that the determinant must equal one in every coordinate system.

The motivation is closely connected to the cosmological-constant problem. In ordinary general relativity, a constant contribution to the matter Lagrangian has stress tensor proportional to $g_{\mu\nu}$ and therefore contributes to the same tensor structure as a cosmological constant. Quantum field theory in curved spacetime allows vacuum contributions from zero-point fluctuations and phase transitions; the discrepancy between naive estimates and the observed late-time curvature scale is the familiar radiative-stability and fine-tuning problem \citep{Weinberg1989,Carroll2001}. UG reorganizes the local gravitational equations so that the trace part is not independently fixed by the metric variation. The resulting integration constant plays the role of the cosmological constant. This reorganization is conceptually important, but it does not by itself constitute a complete solution of the quantum naturalness problem: the treatment of vacuum energy, quantum corrections, and the functional measure remains an active subject \citep{Alvarez2005,AlvarezHerreroValea2013,CarballoRubio2022,Jirousek2023,AlvarezVelascoAja2023}.

The original 1919 proposal was motivated by a question about the role of the metric determinant in the structure of matter; the modern field-theoretic interpretation emerged later. Anderson and Finkelstein formulated a restricted-coordinate theory in which the invariant volume element is held fixed, and subsequent work clarified the relation between the resulting massless spin-two dynamics and the Einstein equations. The 1980s developments by van der Bij, van Dam and Ng, Buchm"uller and Dragon, and Henneaux and Teitelboim established several useful routes from restricted coordinate invariance to the familiar gravitational dynamics with an integration constant \citep{AndersonFinkelstein1971,VanDerBij1982,BuchmullerDragon1988,BuchmullerDragon1989,HenneauxTeitelboim1989}. Unruh's canonical analysis made the role of the unimodular condition in the Hamiltonian description particularly clear \citep{Unruh1989,UnruhWald1989}. The modern literature has since developed classical, canonical, path-integral, and effective-field-theory viewpoints; the 2023 primer by Álvarez and Velasco-Aja provides a unified account of these formulations and their technical details \citep{Alvarez2005,AlvarezGonzalezMartinMartin2016,CarballoRubio2022,Jirousek2023,AlvarezVelascoAja2023}.

The trace-free equations in \cref{eq:tracefree_UG} arise because an allowed metric variation is traceless when the determinant is fixed. Equivalently, one may project the unrestricted metric equations onto their traceless part. This removes the trace equation at the local variational level; the missing scalar equation is recovered as an integration relation after the divergence is taken. The distinction is important: the trace-free tensor equation is local, while the integration constant in \cref{eq:trace_integration} is constant on each connected component and is not fixed by the trace-free equation itself. When a constant vacuum term is included in the matter Lagrangian, its stress tensor is proportional to $g_{\mu\nu}$ and disappears from the traceless projection. The observed curvature scale is then determined by the combination of the integration constant and the vacuum contribution. This is the precise sense in which UG changes the bookkeeping of vacuum energy; it does not automatically remove the radiative-stability question associated with selecting the observed value \citep{Weinberg1989,AlvarezHerreroValea2013,CarballoRubio2022}.

One formulation begins by varying the metric while preserving its determinant. The metric field equations are then trace-free. For minimally coupled matter they take the form
\begin{equation}
R_{\mu\nu}-\frac14 Rg_{\mu\nu}
=\frac{1}{\mpl^2}\left(T_{\mu\nu}-\frac14 Tg_{\mu\nu}\right).
\label{eq:tracefree_UG}
\end{equation}
Taking a covariant divergence and using the contracted Bianchi identity,
$\nabla^\mu R_{\mu\nu}=\frac12\nabla_\nu R$, together with covariant conservation of the matter stress tensor, gives
\begin{equation}
\nabla_\nu\left(R+\frac{T}{\mpl^2}\right)=0.
\end{equation}
On each connected spacetime region this integrates to
\begin{equation}
R+\frac{T}{\mpl^2}=4\Lambda,
\label{eq:trace_integration}
\end{equation}
where $\Lambda$ is a spacetime constant. Substituting this relation back into the trace-free equations recovers the Einstein equations with a cosmological-constant term. The value of $\Lambda$ is an integration constant of the equations, rather than a coupling fixed by the local gravitational action \citep{HenneauxTeitelboim1989,Unruh1989,EllisElstMuruganUzan2011,SaezGomez2016}.

A second presentation fixes the volume form directly at the level of the action. One introduces a Lagrange multiplier $\lambda(x)$ and writes, schematically,
\begin{equation}
S_{\rm U}[g,\Psi,\lambda]
=\int\dd^4x\left[\sqrt{-g}\,\cL(g,\Psi)
-\lambda(x)\big(\sqrt{-g}-\varepsilon_0(x)\big)\right].
\label{eq:UG_lagrange_generic}
\end{equation}
Variation with respect to $\lambda$ imposes the fixed-volume condition. The multiplier contributes to the metric equations in the same algebraic tensor structure as a cosmological-constant term. Its differential equation follows from the divergence of the metric equations and the matter equations. This Lagrange-multiplier description is convenient in scalar--tensor cosmology because the volume constraint is explicit and can be carried through a field redefinition without replacing the theory by a trace-free equation at every step \citep{SaezGomez2016,NojiriOdintsovOikonomou2016,NozariMalekpour2024,MalekpourNozariRajabiRashidi2024}.

A third, fully covariant implementation is the Henneaux--Teitelboim formulation. There the fixed-volume condition is enforced by a vector density (equivalently, a three-form potential), and the cosmological constant appears as the momentum conjugate to the corresponding four-form degree of freedom. This formulation makes the integration-constant interpretation particularly transparent while retaining manifest covariance \citep{HenneauxTeitelboim1989,Unruh1989}. Closely related formulations are described using transverse diffeomorphisms (TDiff), which preserve the prescribed volume form, and Weyl-transverse diffeomorphisms (WTDiff), in which a Weyl symmetry supplements the transverse coordinate symmetry. These formulations illuminate different gauge-theoretic descriptions of unimodular dynamics; their off-shell field variables and gauge structure should be distinguished even when they reproduce the same classical local equations under suitable conditions \citep{BuchmullerDragon1988,BuchmullerDragon1989,AlvarezBlasGarrigaVerdaguer2006,AlvarezVidal2010,AlvarezVelascoAja2023}.

A representative Henneaux--Teitelboim action, in units where $M_{\rm P}=1$, can be written schematically as
\begin{equation}
S_{\rm HT}=\int\dd^4x\left\{\sqrt{-g}\left[\frac12R-\Lambda(x)\right]
+\Lambda(x)\partial_\mu\tau^\mu\right\}+S_{\rm m},
\label{eq:HT_action}
\end{equation}
where $\tau^\mu$ is a vector density. Variation with respect to $\tau^\mu$ gives $\partial_\mu\Lambda=0$, whereas variation with respect to $\Lambda$ relates the metric volume density to the divergence of $\tau^\mu$. The cosmological constant is consequently a canonical/global variable rather than a freely varying local function. A three-form description is equivalent after identifying the vector density with the dual of a three-form potential. This makes explicit how a covariant action can retain the integration-constant character of the cosmological term \citep{HenneauxTeitelboim1989,Unruh1989}.

The restricted-diffeomorphism language provides another useful classification. If a fixed density $\omega(x)$ defines the volume form, the infinitesimal coordinate transformations preserving it obey $\nabla^{(\omega)}_\mu\zeta^\mu=0$. In coordinates with constant $\omega$, this reduces to $\partial_\mu\zeta^\mu=0$. These are transverse diffeomorphisms. In WTDiff models, a local Weyl rescaling is added as a gauge symmetry, and the determinant condition may be viewed as a gauge choice within a larger redundancy. The distinction between a fixed-volume theory and a Weyl-compensated theory is therefore not merely notational: it determines which transformations are gauge symmetries before and after gauge fixing and how the metric determinant is represented off shell \citep{BuchmullerDragon1988,BuchmullerDragon1989,AlvarezBlasGarrigaVerdaguer2006,AlvarezVidal2010,AlvarezVelascoAja2023}.

For the present model the Lagrange-multiplier action is the most direct choice because it keeps the volume constraint as an explicit variational equation. It is also particularly useful for the field redefinition performed below: the multiplier term can be transformed together with the measure, so the determinant constraint is not lost when the gravitational coefficient is normalized. The other formulations are reviewed here to place this choice in the wider UG framework; the numerical calculation itself uses the independent-multiplier action specified in \cref{sec:model}.

For standard UG coupled to covariantly conserved matter, the local gravitational degrees of freedom agree classically with those of general relativity; the integration constant supplies the additional global datum. The equivalence at the level of local classical solutions does not erase the conceptual difference in how the cosmological constant enters the theory, and it does not settle whether the quantum theories have identical measures, radiative corrections, or boundary sectors. These questions have been studied from the perspectives of canonical quantization, BRST symmetry, path integrals, and quantum corrections \citep{Unruh1989,Alvarez2005,AlvarezGonzalezMartinMartin2016,CarballoRubio2022,AlvarezVelascoAja2023}.

The present analysis uses the Lagrange-multiplier formulation in \cref{eq:UG_lagrange_generic}. It is particularly suitable for the inflationary question addressed here: the multiplier is retained as an independent variable during the variational transformation, its on-shell constancy is established in the Einstein frame, and its integration constant is then treated as a parameter of the effective inflationary potential. This allows its effect on the CMB observables to be tested directly.

The Coleman--Weinberg mechanism is a classic example of radiative symmetry breaking. In their 1973 work, Coleman and Weinberg showed that quantum corrections can generate a nontrivial symmetry-breaking vacuum even when the classical scalar theory is scale invariant and contains no explicit mass scale \citep{ColemanWeinberg1973}. In a classically massless gauge--scalar theory, the one-loop effective potential contains logarithms of the background scalar field. The renormalization scale and the running couplings combine so that a dimensionful vacuum expectation value emerges from dimensionless parameters. This phenomenon is called dimensional transmutation.

The schematic one-loop structure is
\begin{equation}
V_{\rm 1-loop}(\phi)\sim \phi^4\left[c_0+c_1\ln\left(\frac{\phi}{\mu}\right)\right],
\end{equation}
where the coefficients depend on the particle content and couplings of the underlying quantum field theory. More generally, in a renormalizable theory with background-dependent masses $m_i(\phi)$, the one-loop contribution has the form
\begin{equation}
V_1(\phi)=\frac{1}{64\pi^2}\sum_i(-1)^{F_i}n_i m_i^4(\phi)
\left[\ln\frac{m_i^2(\phi)}{\mu^2}-c_i\right],
\label{eq:one_loop_general}
\end{equation}
where $F_i$ distinguishes bosons and fermions, $n_i$ counts their degrees of freedom, $\mu$ is the renormalization scale, and $c_i$ depends on the renormalization prescription. Gauge boson, scalar, and fermion loops therefore contribute with different signs and multiplicities. In the original scalar electrodynamics example, gauge-field fluctuations generate the logarithmic term that can destabilize the origin and produce a nonzero minimum even when the classical scalar mass is zero \citep{ColemanWeinberg1973}. The logarithm also signals the dependence of the effective description on the renormalization scale. Renormalization-group improvement resums the leading logarithms by allowing couplings to run with the scale; choosing a scale of order the background field is a common way to keep the logarithms moderate along the inflationary trajectory.

After choosing renormalization conditions at the radiative minimum $\phi_0$, a convenient effective potential for the real scalar inflaton is
\begin{equation}
V_{\CW}(\phi)=\gamma\left[\phi^4\left(\ln\frac{\phi}{\phi_0}-\frac14\right)+\frac{\phi_0^4}{4}\right],\qquad \gamma>0.
\label{eq:CW_potential}
\end{equation}
The constant term is chosen so that the potential vanishes at the symmetry-breaking vacuum. The renormalization conditions make $\phi_0$ a zero-energy stationary point of the potential; for $\gamma>0$, the curvature there is positive, so it is the radiative minimum.

CW-type potentials were incorporated into inflationary model building soon after the development of new inflation. The radiatively generated flat region near the origin motivates the small-field or hilltop branch, while the portion above the symmetry-breaking scale provides a distinct large-field branch \citep{AlbrechtSteinhardt1982,ShafiVilenkin1984,SenoguzShafi2006}. Subsequent analyses examined the consistency of CW inflation with the increasingly precise CMB measurements and its embedding into particle-physics models, including grand-unified and classically scale-invariant extensions \citep{RehmanShafiWickman2008,BarenboimChunLee2014,KannikeRacioppiRaidal2014}. The two branches have different observational challenges. Near the origin, the potential is sufficiently flat that the tensor amplitude is strongly suppressed, but the scalar tilt can be too red. Above the minimum, the slope is steeper and the tensor amplitude is appreciable, so suppressing $r$ while preserving an acceptable $\ns$ becomes important.

The potential in \cref{eq:CW_potential} is a renormalized effective potential, not a generic quartic monomial. Its logarithmic derivative controls the slow-roll trajectory and distinguishes it from the simple $\lambda\phi^4$ model. The coefficient $\gamma$ fixes the overall energy scale, whereas the shape is controlled by $\phi_0$ and by the dimensionless field value in units of the reduced Planck mass. The scalar amplitude measured by the CMB is much smaller than unity, which typically selects a very small radiative coupling when the inflationary field range is Planckian. This is consistent with the role of $\gamma$ as a radiative coupling rather than a freely adjustable order-one parameter \citep{ShafiVilenkin1984,SenoguzShafi2006,BarenboimChunLee2014}.

The two inflationary branches follow directly from the shape of the same symmetry-breaking potential. For $0<\phi<\phi_0$, the field moves on the inner side of the radiative minimum; the potential is very flat near the origin, so the tensor amplitude is small, while the curvature of the potential can make the scalar tilt too red. For $\phi>\phi_0$, the field evolves on the outer side of the minimum and the slope is larger, yielding a larger tensor signal. The two branches are therefore not two different potentials but two distinct inflationary trajectories of a single effective potential. Their observational behavior has been studied in CW and related symmetry-breaking inflation models as CMB constraints have improved \citep{ShafiVilenkin1984,SenoguzShafi2006,RehmanShafiWickman2008,BarenboimChunLee2014,Bostan2020,BostanRoyChoudhury2024}.

The normalization in \cref{eq:CW_potential} is useful for the present analysis because it sets both the vacuum energy and the first derivative to zero at $\phi_0$, while retaining a positive curvature there
which identifies $\phi_0$ as a local minimum for positive $\gamma$. During inflation the field is displaced from this vacuum, and it is precisely the logarithmic term that controls the slope over the relevant field range. This connection between radiative symmetry breaking, the vacuum expectation value, and the inflationary trajectory motivates normalizing the nonminimal coupling at the same field value.

A scalar field in curved spacetime admits a four dimensional interaction proportional to $\phi^2R$. Such interactions are generated by renormalization in curved spacetime and have long been used to modify the inflationary dynamics. In Higgs inflation, for example, a large nonminimal coupling flattens the effective potential at large field values and can lower the predicted tensor amplitude \citep{FakirUnruh1990,BezrukovShaposhnikov2008}. For a scalar with a nonzero vacuum expectation value, however, the distinguished point in field space is not the origin but the symmetry-breaking vacuum $\phi_0$. It is therefore natural to normalize the curvature coupling at that vacuum and use
\begin{equation}
F(\phi)=1+\xi(\phi^2-\phi_0^2),\qquad F(\phi_0)=1.
\label{eq:F_shifted}
\end{equation}
The gravitational action is written as $\frac12F(\phi)R$ in reduced Planck units.

The shift has several motivations. First, it makes the coefficient of the Einstein--Hilbert term equal to its reference value after the inflaton settles at the radiative minimum. In this sense the effective squared Planck mass, $M_{\rm eff}^2(\phi)=M_{\rm P}^2F(\phi)$, returns to $M_{\rm P}^2$ at $\phi=\phi_0$. Second, it ties the nonminimal interaction to the physical vacuum selected by the CW potential and avoids leaving a constant contribution $\xi\phi_0^2$ in the normalization of the gravitational term at that vacuum. Third, it preserves the usual quadratic field dependence away from the minimum, so the coupling can still reshape the inflationary potential during the large- or small-field phase. This is a physically motivated normalization condition rather than a unique consequence of the CW mechanism.

The same vacuum-shifted function has been used in nonminimally coupled CW analyses. In particular, the Palatini study of Bostan employs $F(\phi)=1+\xi(\phi^2-v^2)$ for a scalar with nonzero vacuum expectation value, and the later analysis by Bostan and Roy Choudhury uses the same form for the CW potential \citep{Bostan2020,BostanRoyChoudhury2024}. These studies motivate the functional form but do not determine the numerical predictions in the present metric theory. We adopt the shifted coupling in the metric formulation. This distinction matters: the metric formulation generates the additional kinetic contribution proportional to $3(F_{,\phi}/F)^2/2$ after the Weyl transformation, whereas in the Palatini formulation the independent connection changes the canonical normalization. We therefore derive the kinetic factor and all inflationary expressions from the metric action used here.

There is also a useful effective-field-theory interpretation. The function $F$ is the coefficient of the two-derivative curvature term and hence determines the field-dependent gravitational normalization. Requiring $F(\phi_0)=1$ fixes this normalization at the vacuum in which the CW potential is minimized. The shift does not eliminate nonminimal effects during inflation: for $\phi\neq\phi_0$, $F$ differs from unity and changes both the Einstein-frame potential through $F^{-2}$ and the kinetic geometry through $K(\phi)$. The same parameter $\xi$ therefore acts in two linked ways, changing the potential height and slope as well as the canonical distance traversed by the scalar. This is why optimizing $\xi$ is not equivalent to multiplying the original CW potential by a simple constant factor.

Throughout the numerical analysis we require $F(\phi)>0$ and $K(\phi)>0$, so that the effective gravitational coupling is positive and the scalar degree of freedom has a positive kinetic term.

The central question is whether the unimodular integration constant can improve the observational status of a CW model with the vacuum-shifted nonminimal coupling. We focus on the effect of this constant rather than suppressing it by setting it to zero. The analysis is organized as follows. In \cref{sec:model} we specify the fixed-volume scalar--tensor action. In \cref{sec:einstein} we transform the action to the Einstein frame, derive the constancy of the multiplier using the Einstein-frame equations, and discuss the metric and scalar field redefinitions. In \cref{sec:slowroll} we develop the potential slow-roll formalism. The numerical workflow and posterior inputs are given in \cref{sec:numerics}; the large-field results and the allowed intervals are presented in \cref{sec:large}. Finally, \cref{sec:small} summarizes the small-field branch, followed by the conclusions.

\section{Unimodular Coleman--Weinberg model}
\label{sec:model}

We work with metric signature $(-,+,+,+)$ and use reduced Planck units, $M_{\rm P}=1$, unless stated otherwise. The scalar--tensor action that defines the model is
\begin{equation}
S=\int\dd^4x\left\{\sqrt{-g}\left[\frac12F(\phi)R-\frac12g^{\mu\nu}\partial_\mu\phi\partial_\nu\phi-V_{\CW}(\phi)\right]
-\lambda(x)\left(\sqrt{-g}-1\right)\right\}.
\label{eq:starting_action}
\end{equation}
Here $F(\phi)$ is given by \cref{eq:F_shifted}, $V_{\CW}$ by \cref{eq:CW_potential}, and $\lambda(x)$ is an the Lagrange multiplier, treated as an independent variational variable (varied separately from the metric and scalar field). It has no kinetic term and is not a propagating degree of freedom. The normalization $\sqrt{-g}=1$ is the unimodular coordinate choice for the fixed volume density. The multiplier is not prescribed as a function of $\phi$ before variation.

The multiplier equation is
\begin{equation}
\frac{\delta S}{\delta\lambda}=0
\quad\Longrightarrow\quad \sqrt{-g}=1.
\label{eq:det_constraint}
\end{equation}
The role of this condition in the present paper is to define the constrained gravitational system. We will carry it through the Weyl transformation and use the complete Einstein-frame variational equations to identify the on-shell multiplier.

\section{Transformation to the Einstein frame}
\label{sec:einstein}

\subsection{Weyl transformation of the metric}

We define the Einstein-frame metric by
\begin{equation}
g^E_{\mu\nu}=F(\phi)g_{\mu\nu},\qquad F(\phi)>0.
\label{eq:weyl}
\end{equation}
The inverse metric and volume element satisfy
\begin{equation}
g_E^{\mu\nu}=F^{-1}g^{\mu\nu},\qquad \sqrt{-g}=F^{-2}\sqrt{-g_E}.
\label{eq:measure_transform}
\end{equation}
In four spacetime dimensions the Ricci scalar transforms as
\begin{equation}
R=F\left[R_E+3\Box_E\ln F-\frac32g_E^{\mu\nu}\partial_\mu\ln F\partial_\nu\ln F\right].
\label{eq:ricci_transform}
\end{equation}
After inserting these relations into \cref{eq:starting_action}, discarding the total derivative proportional to $\Box_E\ln F$, and retaining the multiplier as an independent variational variable, the action becomes
\begin{align}
S_E={}&\int\dd^4x\sqrt{-g_E}\left[\frac12R_E-\frac12K(\phi)(\nabla_E\phi)^2-\frac{V_{\CW}(\phi)}{F(\phi)^2}\right]\notag\\
&-\int\dd^4x\,\lambda(x)\left[F(\phi)^{-2}\sqrt{-g_E}-1\right],
\label{eq:EF_action_unreduced}
\end{align}
where
\begin{equation}
K(\phi)=\frac1{F(\phi)}+\frac32\left(\frac{F_{,\phi}}{F}\right)^2.
\label{eq:K_general}
\end{equation}
The transformed unimodular constraint is
\begin{equation}
F(\phi)^{-2}\sqrt{-g_E}=1.
\label{eq:EF_unimodular_constraint}
\end{equation}
It is important that the fixed-volume condition transforms along with the metric. In the Einstein frame it is not simply $\sqrt{-g_E}=1$; rather, the factor $F^{-2}$ is part of the transformed volume constraint.

\subsection{Constancy of the multiplier derived in the Einstein frame}
\label{subsec:lambda_constant_E}

For the derivation it is convenient to combine the two terms proportional to $\sqrt{-g_E}$ and define
\begin{equation}
U(\phi,\lambda)=\frac{V_{\CW}(\phi)+\lambda}{F(\phi)^2}.
\label{eq:U_phi_lambda}
\end{equation}
The action can then be written as
\begin{equation}
S_E=\int\dd^4x\sqrt{-g_E}\left[\frac12R_E-\frac12K(\phi)(\nabla_E\phi)^2-U(\phi,\lambda)\right]
+\int\dd^4x\,\lambda(x).
\label{eq:EF_action_U}
\end{equation}
The final coordinate-volume term does not depend on the Einstein metric or on $\phi$, but it is essential in the multiplier variation. Variation with respect to $\lambda$ reproduces \cref{eq:EF_unimodular_constraint}.

The metric equation obtained from \cref{eq:EF_action_U} is
\begin{equation}
G^E_{\mu\nu}=T^{(\phi,\lambda)}_{\mu\nu},
\label{eq:EF_Einstein_eq}
\end{equation}
with
\begin{equation}
T^{(\phi,\lambda)}_{\mu\nu}
=K\partial_\mu\phi\partial_\nu\phi
-g^E_{\mu\nu}\left[\frac12K(\nabla_E\phi)^2+U(\phi,\lambda)\right].
\label{eq:EF_stress}
\end{equation}
The scalar equation is obtained by varying $\phi$ while holding the multiplier $\lambda$ fixed:
\begin{equation}
\nabla^E_\mu\left(K\nabla_E^\mu\phi\right)
-\frac12K_{,\phi}(\nabla_E\phi)^2
-U_{,\phi}\big|_{\lambda}=0.
\label{eq:EF_scalar_eq}
\end{equation}
Taking the covariant divergence of \cref{eq:EF_stress} gives the identity
\begin{align}
\nabla_E^\mu T^{(\phi,\lambda)}_{\mu\nu}
={}&\left[\nabla^E_\mu(K\nabla_E^\mu\phi)
-\frac12K_{,\phi}(\nabla_E\phi)^2-U_{,\phi}\big|_{\lambda}\right]\partial_\nu\phi\notag\\
&-U_{,\lambda}\,\partial_\nu\lambda.
\label{eq:stress_divergence}
\end{align}
The contracted Bianchi identity gives $\nabla_E^\mu G^E_{\mu\nu}=0$. Using the Einstein equation, the scalar equation, and \cref{eq:stress_divergence}, we obtain
\begin{equation}
U_{,\lambda}\,\partial_\nu\lambda=0.
\label{eq:lambda_gradient_intermediate}
\end{equation}
But from \cref{eq:U_phi_lambda},
\begin{equation}
U_{,\lambda}=F(\phi)^{-2}.
\end{equation}
Since the conformal transformation is regular and $F>0$, this coefficient is nonzero. It follows that
\begin{equation}
\partial_\nu\lambda=0,\qquad \lambda(x)=\lambdat=\text{constant}.
\label{eq:lambda_constant_E}
\end{equation}
Thus the multiplier is an integration constant of the Einstein-frame field equations. This conclusion follows from the independent variation of the multiplier, the scalar Euler--Lagrange equation, and the Bianchi identity; it is not imposed as a slow-roll condition. The result allows us to replace $\lambda$ by $\lambdat$ in the effective potential.

\subsection{Canonical scalar field in the Einstein frame}
\label{subsec:scalar_field_redefinition}

The conformal transformation first changes the metric according to \cref{eq:weyl}. In carrying this out, the scalar field $\phi$ is retained as the field-space variable in which the original potential and coupling are written. This does not yet make its kinetic term canonical. In fact, two contributions enter the Einstein-frame kinetic sector: the original scalar kinetic term is rescaled by the transformed inverse metric and volume element, and the conformal transformation of the curvature term generates an additional derivative contribution. Combining them gives the coefficient
\begin{equation}
K(\phi)=\frac{1}{F(\phi)}+\frac{3}{2}\left(\frac{F_{,\phi}}{F}\right)^2,
\end{equation}
as obtained in \cref{eq:K_general}. This is the general scalar--tensor procedure described by Fujii and Maeda \citep{FujiiMaeda2009} such that the conformal factor is chosen to put the curvature term in Einstein--Hilbert form, after which the contributions to the scalar kinetic term are collected \citep{FaraoniGunzigNardone1999,Kaiser2010}.

A new scalar coordinate is then defined so that the combined kinetic term becomes canonical. Provided $K(\phi)>0$, we introduce
\begin{equation}
\frac{\dd\chi}{\dd\phi}=\sqrt{K(\phi)},\qquad
\chi(\phi)=\int^{\phi}\sqrt{K(\varphi)}\,\dd\varphi.
\label{eq:canonical_map}
\end{equation}
Consequently, $(\nabla_E\chi)^2=K(\phi)(\nabla_E\phi)^2$, and the Einstein-frame action takes the canonical scalar form
\begin{equation}
S_E=\int\dd^4x\sqrt{-g_E}\left[\frac12R_E-\frac12(\nabla_E\chi)^2-U_{\eff}(\chi)\right]
+\lambdat\int\dd^4x,
\label{eq:EF_canonical_action}
\end{equation}
where $U_{\eff}(\chi)=U_{\eff}(\phi(\chi))$. The map $\phi\mapsto\chi$ is an invertible reparametrization of the one-dimensional scalar-field space wherever $K>0$; it introduces no additional degree of freedom. Thus the Einstein-frame gravitational sector is minimally coupled, while the scalar kinetic term becomes canonical only after this second field redefinition. In the numerical calculation we retain $\phi$ as the integration variable and include $K(\phi)$ explicitly, which is equivalent to using $\chi$ when the map is monotonic.
\subsection{Einstein-frame potential for the shifted CW model}

Using \cref{eq:lambda_constant_E}, the effective potential is
\begin{equation}
U_{\eff}(\phi)=\frac{V_{\CW}(\phi)+\lambdat}{F(\phi)^2}.
\label{eq:Ueff}
\end{equation}
For the vacuum-shifted coupling case we have $F_{,\phi}=2\xi\phi$ and $F_{,\phi\phi}=2\xi$.
Therefore
\begin{equation}
K(\phi)=\frac{1}{F}+\frac{6\xi^2\phi^2}{F^2}
=\frac{F+6\xi^2\phi^2}{F^2}.
\label{eq:K_specific}
\end{equation}
It is useful to define
\begin{equation}
E(\phi)=F(\phi)+\frac32F_{,\phi}^2
=1+\xi(\phi^2-\phi_0^2)+6\xi^2\phi^2,
\end{equation}
so that $K=E/F^2$ and $\dd\chi/\dd\phi=\sqrt{E}/F$. The derivative of the effective potential is
\begin{equation}
U_{\eff,\phi}
=\frac{F V_{\CW,\phi}-2F_{,\phi}(V_{\CW}+\lambdat)}{F^3}.
\label{eq:Ueff_prime}
\end{equation}
The second derivative, required for $\eta_V$, is obtained by differentiating \cref{eq:Ueff_prime} with respect to $\phi$. For clarity, the canonical derivatives are related to derivatives in the numerical field coordinate by
\begin{equation}
U_{,\chi}=\frac{U_{\eff,\phi}}{\sqrt K},\qquad
U_{,\chi\chi}=\frac1K\left(U_{\eff,\phi\phi}-\frac{K_{,\phi}}{2K}U_{\eff,\phi}\right).
\label{eq:canonical_derivatives}
\end{equation}

\section{Potential slow-roll inflation}
\label{sec:slowroll}

In the Einstein frame, the gravitational sector is the standard Einstein--Hilbert term and the scalar is canonical when expressed in $\chi$. For a spatially flat Friedmann--Lema\^itre--Robertson--Walker background, the exact homogeneous equations are
\begin{align}
3H_E^2&=\frac12\dot\chi^2+U_{\eff}(\chi),\label{eq:Friedmann_E}\\
\dot H_E&=-\frac12\dot\chi^2,\label{eq:Raychaudhuri_E}\\
\ddot\chi+3H_E\dot\chi+U_{,\chi}&=0.
\label{eq:KleinGordon_E}
\end{align}
During slow roll, the kinetic energy and acceleration of the inflaton are small compared with the potential and Hubble-friction terms, respectively. The leading equations become
\begin{equation}
3H_E^2\simeq U_{\eff},\qquad 3H_E\dot\chi\simeq-U_{,\chi}.
\label{eq:slowroll_eqs}
\end{equation}
Here, ``first order'' refers to the leading-order slow-roll approximation used to obtain the background and spectral observables; we do not solve the perturbation mode equations numerically. The scalar and tensor spectra are evaluated with the standard leading-order potential slow-roll formulae.

The potential slow-roll parameters are
\begin{equation}
\epsilon_V=\frac12\left(\frac{U_{,\chi}}{U_{\eff}}\right)^2,
\qquad
\eta_V=\frac{U_{,\chi\chi}}{U_{\eff}}.
\label{eq:slowroll_params_chi}
\end{equation}
Inflation proceeds while $\epsilon_V<1$ and $|\eta_V|<1$; the end of inflation is defined by the first point on the selected branch at which
\begin{equation}
\epsilon_V(\phi_{\rm end})=1.
\label{eq:end_inflation}
\end{equation}
Using $\phi$ as the numerical variable, the same slow-roll parameters are
\begin{align}
\epsilon_V(\phi)&=\frac{1}{2K(\phi)}\left(\frac{U_{\eff,\phi}}{U_{\eff}}\right)^2,
\label{eq:eps_phi}\\
\eta_V(\phi)&=\frac{1}{K(\phi)U_{\eff}}
\left[U_{\eff,\phi\phi}-\frac{K_{,\phi}}{2K}U_{\eff,\phi}\right].
\label{eq:eta_phi}
\end{align}
The number of e-folds between horizon exit and the end of inflation is
\begin{equation}
N=\int_{t_*}^{t_{\rm end}}H_E\dd t
\simeq\int_{\phi_{\rm end}}^{\phi_*}
K(\phi)\frac{U_{\eff}(\phi)}{U_{\eff,\phi}(\phi)}\,\dd\phi.
\label{eq:efolds}
\end{equation}
This oriented integral is positive on both branches when the field rolls monotonically from $\phi_*$ to $\phi_{\rm end}$.

At first order in the potential slow-roll expansion, the scalar and tensor spectra at horizon crossing are
\begin{equation}
\Prm(k_*)\simeq\frac{U_{\eff}(\phi_*)}{24\pi^2\epsilon_V(\phi_*)},
\qquad
\Pt(k_*)\simeq\frac{2U_{\eff}(\phi_*)}{3\pi^2},
\label{eq:spectra}
\end{equation}
which give
\begin{equation}
\As\simeq\frac{U_{\eff}(\phi_*)}{24\pi^2\epsilon_V(\phi_*)},\qquad
\ns\simeq1-6\epsilon_{V*}+2\eta_{V*},\qquad
r\simeq16\epsilon_{V*}.
\label{eq:observables}
\end{equation}
We impose the observed scalar amplitude $\As=2.1\times10^{-9}$ when normalizing the model. The slow-roll formalism and the conversion between the potential derivatives and the observables follow the standard single-field treatment \citep{LiddleParsonsBarrow1994,StewartLyth1993,LiddleLeach2003}. The numerical calculation below uses these leading-order expressions throughout.

\section{Numerical strategy and observational inputs}
\label{sec:numerics}

\subsection{Conventions and branch definitions}

The numerical calculations use
\begin{equation}
M_{\rm P}=1,\qquad \phi_0=1,\qquad \As=2.1\times10^{-9},\qquad N\in\{50,60,70\}.
\label{eq:numerical_conventions}
\end{equation}
All scalar-field values are expressed in reduced-Planck units: the dimensionless field used numerically is the dimensionful field divided by $M_{\rm P}$. Thus $\phi_0=1$ means that the vacuum expectation value is one reduced Planck mass, not that a dimensionful field has been set equal to a unitless number. A star on a field variable, such as $\phi_*$, denotes its value at horizon exit of the pivot scale; ``end'' denotes the end of inflation.
The large-field branch is defined by
\begin{equation}
\phi_* > \phi_{\rm end}>\phi_0,
\label{eq:large_branch}
\end{equation}
with the field rolling toward the symmetry-breaking vacuum. The small-field branch is defined by $0<\phi_*<\phi_{\rm end}<\phi_0$ and is discussed separately in \cref{sec:small}.

For each choice of $(N,\xi,\lambdat)$, a physically admissible numerical point must satisfy four conditions simultaneously: the first end-of-inflation crossing is found from $\epsilon_V=1$; the e-fold integral equals the selected value of $N$; the scalar amplitude equals $2.1\times10^{-9}$ after the normalization parameter $\gamma$ is determined; and the trajectory remains on the selected branch with $F>0$ and $K>0$. The multiplier feasibility interval is the connected range in which all of these conditions have a solution with positive finite $\gamma$.

The multiplier scan also gives the response of the two inflationary observables. The scalar-index and tensor-to-scalar-ratio trajectories are presented as separate plots in Figure~3(a) and Figure~3(b), respectively. They use the same amplitude-normalized solutions and the same case-dependent feasibility domain as the normalization calculation. The combined $\gamma(\lambdat)$ plot is Figure~4; it displays the three e-fold cases together and marks each case's own positive and negative feasibility ceilings.

\subsection{How the preferred \texorpdfstring{$\xi$}{xi} is selected}
\label{subsec:xi_selection}

The nonminimal coupling is selected separately for each e-fold number rather than assigning one representative value to all trajectories. The selection is performed first on the zero-multiplier slice, $\lambdat=0$, so that the influence of $\xi$ can be isolated from the additional integration constant. For every fixed $N$, the numerical procedure is:
\begin{enumerate}
\item Choose a trial $\xi$ in the scan interval $10^{-3}\leq\xi\leq10^{-1}$.
\item Determine $\phi_{\rm end}$ from $\epsilon_V(\phi_{\rm end})=1$.
\item Solve \cref{eq:efolds} for $\phi_*$ at the chosen $N$.
\item Evaluate $\ns$ and $r$ at horizon crossing and store the result as a point on the $\xi$ trajectory.
\item Refine the minimum of $r(\xi)$ continuously in $\log_{10}\xi$ rather than selecting a coarse-grid point.
\end{enumerate}
At $\lambdat=0$, the effective potential and all its field derivatives are proportional to $\gamma$. The ratios entering $\epsilon_V$, $\eta_V$, and the e-fold integral are therefore independent of the overall normalization $\gamma$. The minimization of $r(\xi)$ is consequently a calculation of the shape of the potential and can be performed before amplitude normalization. Once $\xi_{\rm best}$ is found, $\gamma$ is fixed by \cref{eq:observables}.

The preferred values, quoted to the $10^{-5}$ precision requested for the numerical presentation, are
\begin{equation}
\xi_{\rm best}(50)=0.05478,\qquad
\xi_{\rm best}(60)=0.04609,\qquad
\xi_{\rm best}(70)=0.03981.
\label{eq:xi_best_values}
\end{equation}
The values are minima of $r$ on the $\lambdat=0$ slice within the stated scan interval. This prescription defines a reproducible baseline for testing the independent effect of the unimodular integration constant.
\FloatBarrier
\begin{figure}[H]
\centering
\includegraphics[width=0.82\textwidth]{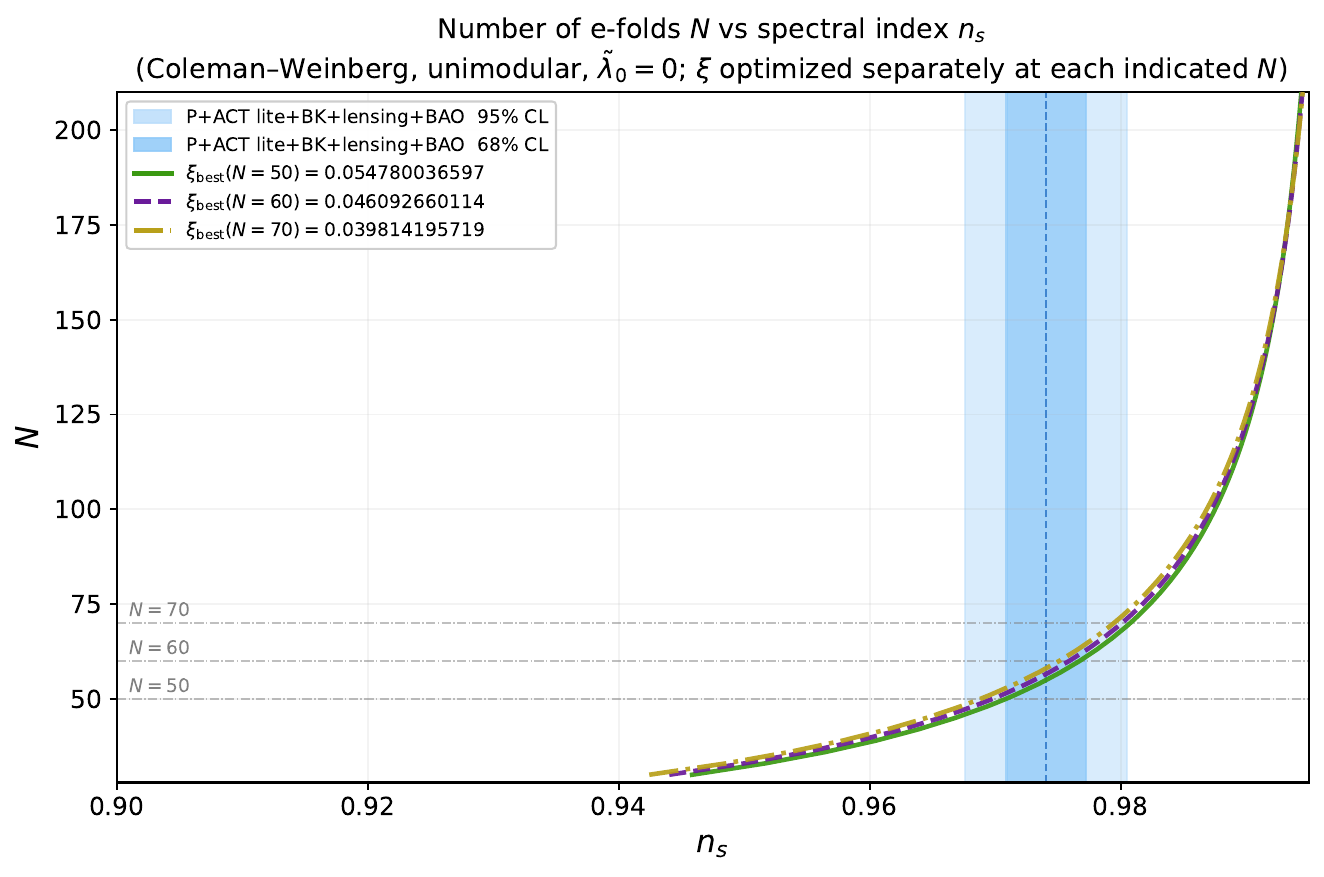}
\caption{The large-field relation between the e-fold number $N$ and the scalar spectral index $\ns$ at $\lambdat=0$, with $\xi$ optimized separately for $N=50$, $60$, and $70$. The observational band is that of the combined Planck+ACT lite+BK+lensing+BAO posterior samples.}
\label{fig:large_N_ns}
\end{figure}
\FloatBarrier
\begin{figure}[H]
\centering
\includegraphics[width=0.86\textwidth]{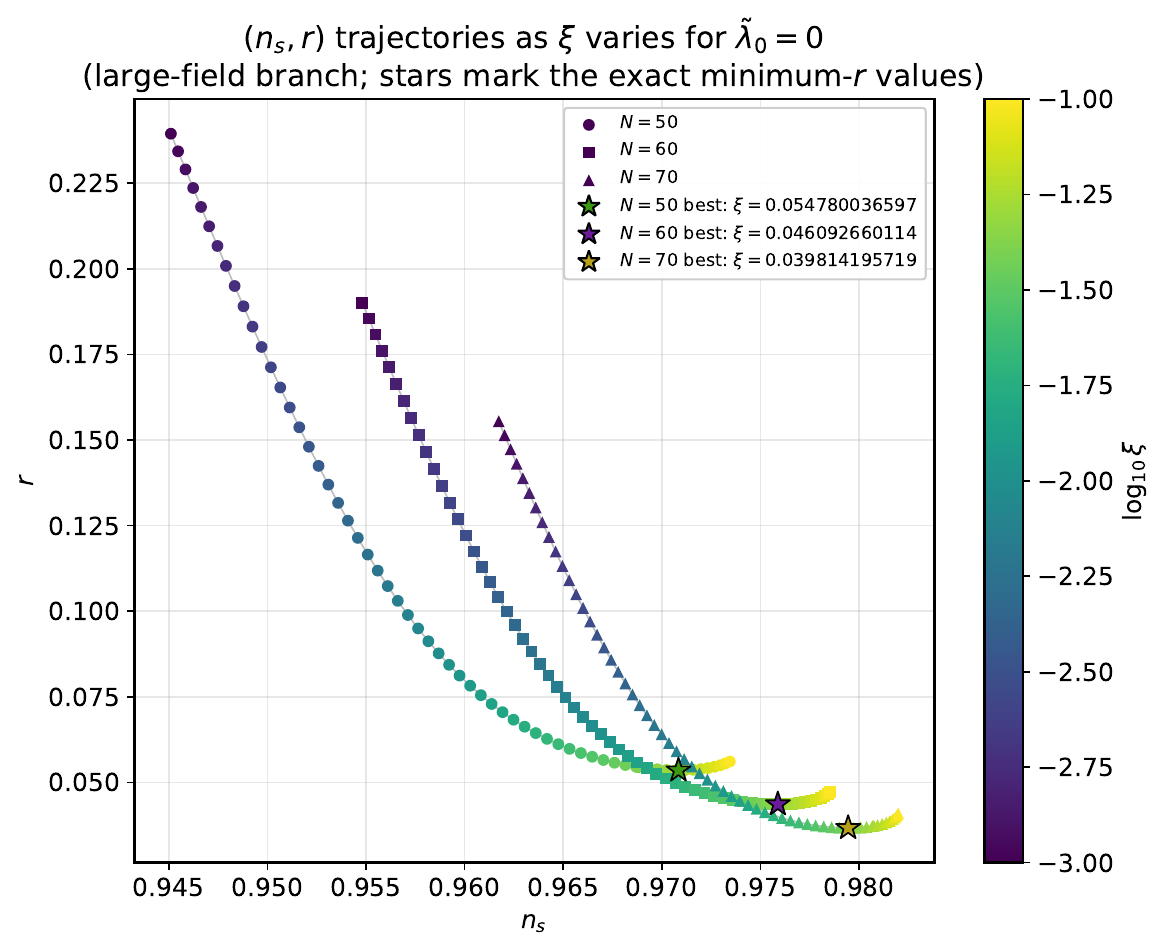}
\caption{Large-field trajectories in the $(\ns,r)$ plane as $\xi$ varies at $\lambdat=0$ for $N=50$, $60$, and $70$. The color scale represents $\log_{10}\xi$. The stars mark the minimum-$r$ points used to define the preferred values in \cref{eq:xi_best_values}. The three trajectories demonstrate why $\xi$ is optimized separately for each e-fold number.}
\label{fig:xi_scan}
\end{figure}

\subsection{Self-consistent normalization and the allowed multiplier interval}
\label{subsec:lambda_interval}

After fixing $\xi=\xi_{\rm best}(N)$, we vary $\lambdat$ and solve the amplitude normalization anew at every point. The reason is structural: for nonzero $\lambdat$, the term $\lambdat/F^2$ is not proportional to $\gamma$. It changes the shape of $U_{\eff}$ and therefore changes $\epsilon_V$, $\eta_V$, the end point, and the horizon-crossing field. It is not consistent to vary $\lambdat$ while holding $\gamma$, $\phi_{\rm end}$, and $\phi_*$ at their zero-multiplier values.

The self-consistent procedure is as follows:
\begin{enumerate}
\item Fix $N$ and the corresponding $\xi_{\rm best}(N)$, and choose a trial positive or negative $\lambdat$.
\item For a trial $\gamma$, construct $U_{\eff}$ and its derivatives, and find the first large-field solution of $\epsilon_V=1$.
\item With that end point, solve the e-fold equation for $\phi_*$ at the selected $N$.
\item Evaluate $\As$ at horizon crossing and vary $\gamma$ until $\As=2.1\times10^{-9}$.
\item Recompute and store $\ns$, $r$, $\gamma$, $\phi_*$, and $\phi_{\rm end}$ for the normalized solution.
\end{enumerate}
The feasibility ceiling on either sign of $\lambdat$ is the endpoint of the connected interval for which this full system has a valid large-field solution. It is not imposed as a prior on the multiplier and is not defined by an observational cut. It follows from the simultaneous requirements of the end condition, the e-fold condition, positivity of the conformal and kinetic factors, and the scalar-amplitude normalization. In particular, outside this interval no positive finite value of $\gamma$ maintains the prescribed amplitude together with the selected inflationary branch.

The resulting scan is strongly asymmetric in sign. The negative branch extends to much larger absolute values than the positive branch. The numerical terminal values are
\begin{center}
\begin{tabular}{c c c}
\toprule
$N$ & $\widetilde{\lambda}_{0,\min}$ & $\widetilde{\lambda}_{0,\max}$\\
\midrule
50 & $-4.3242\times10^{-3}$ & $9.4552\times10^{-12}$\\
60 & $-2.6270\times10^{-3}$ & $5.8449\times10^{-12}$\\
70 & $-1.7109\times10^{-3}$ & $3.8510\times10^{-12}$\\
\bottomrule
\end{tabular}
\end{center}
These endpoints characterize the amplitude-normalized model family. Their very different magnitudes on the two sides of zero are central to the analysis: positive and negative integration constants do not generate symmetric deformations of the CW potential.

The normalized spectral-index and tensor-to-scalar-ratio responses are presented separately in Figure~3(a) and Figure~3(b). Both plots use the same case-dependent feasibility intervals and the same self-consistent solutions described above. They display how the multiplier changes the shape of the effective potential and, through it, the two principal inflationary observables.

The $\gamma(\lambdat)$ relation is shown in \cref{fig:gamma_lambda}. The figure displays all three e-fold cases in one panel and marks each case's own positive and negative feasibility ceiling. Thus, the plotted endpoints are determined by the numerical solution itself; zero is not used as a common upper cutoff.

\begin{figure}[H]
\centering
\includegraphics[width=0.90\textwidth]{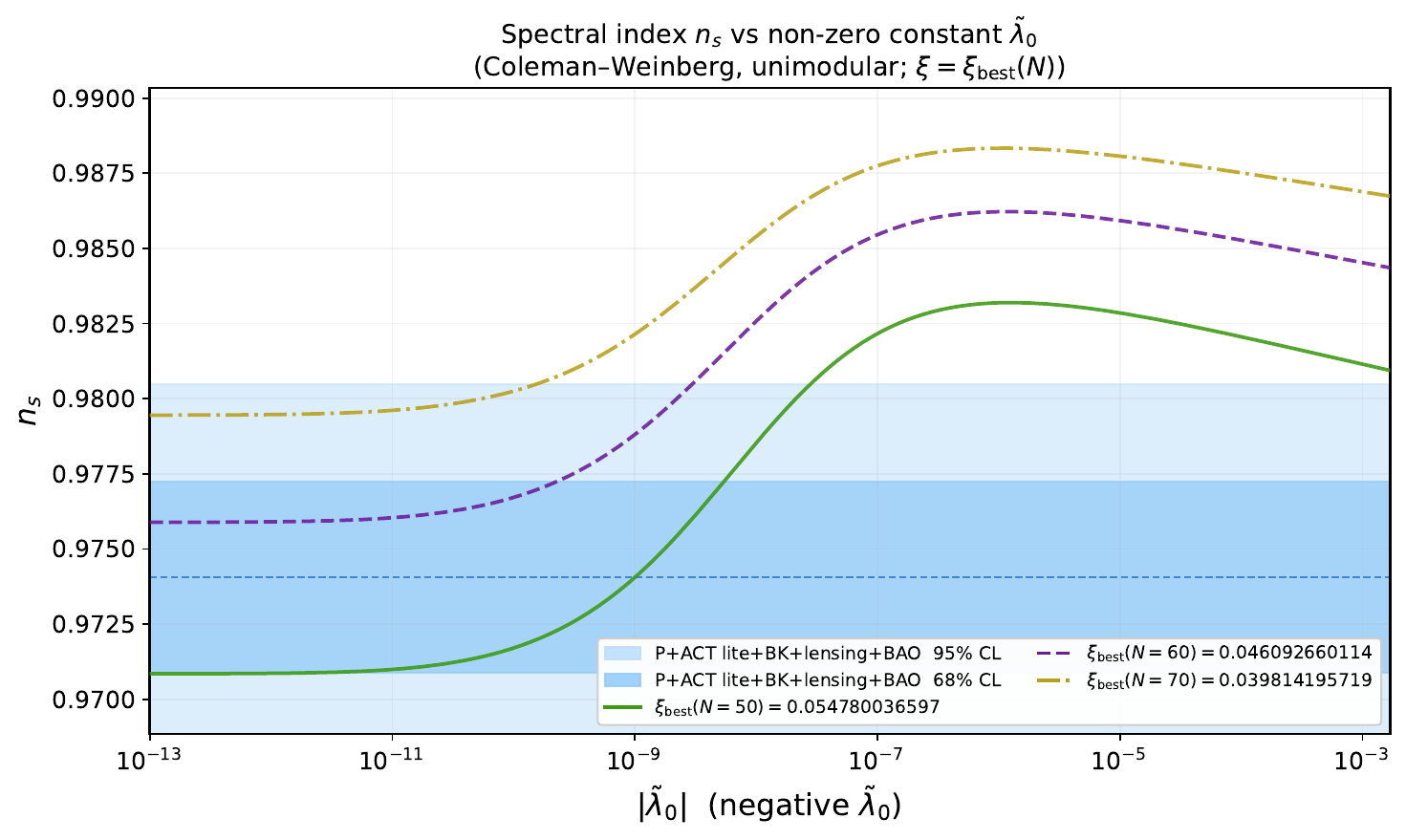}
\caption{(a) Scalar spectral index $\ns$ as a function of the negative unimodular integration constant for the large-field branch, evaluated at $\xi=\xi_{\rm best}(N)$. Each point is independently normalized to $\As=2.1\times10^{-9}$ and satisfies the selected e-fold number.}
\label{fig:lambda_ns}
\end{figure}

\FloatBarrier

\begin{figure}[tbp]
\ContinuedFloat
\centering
\includegraphics[width=0.90\textwidth]{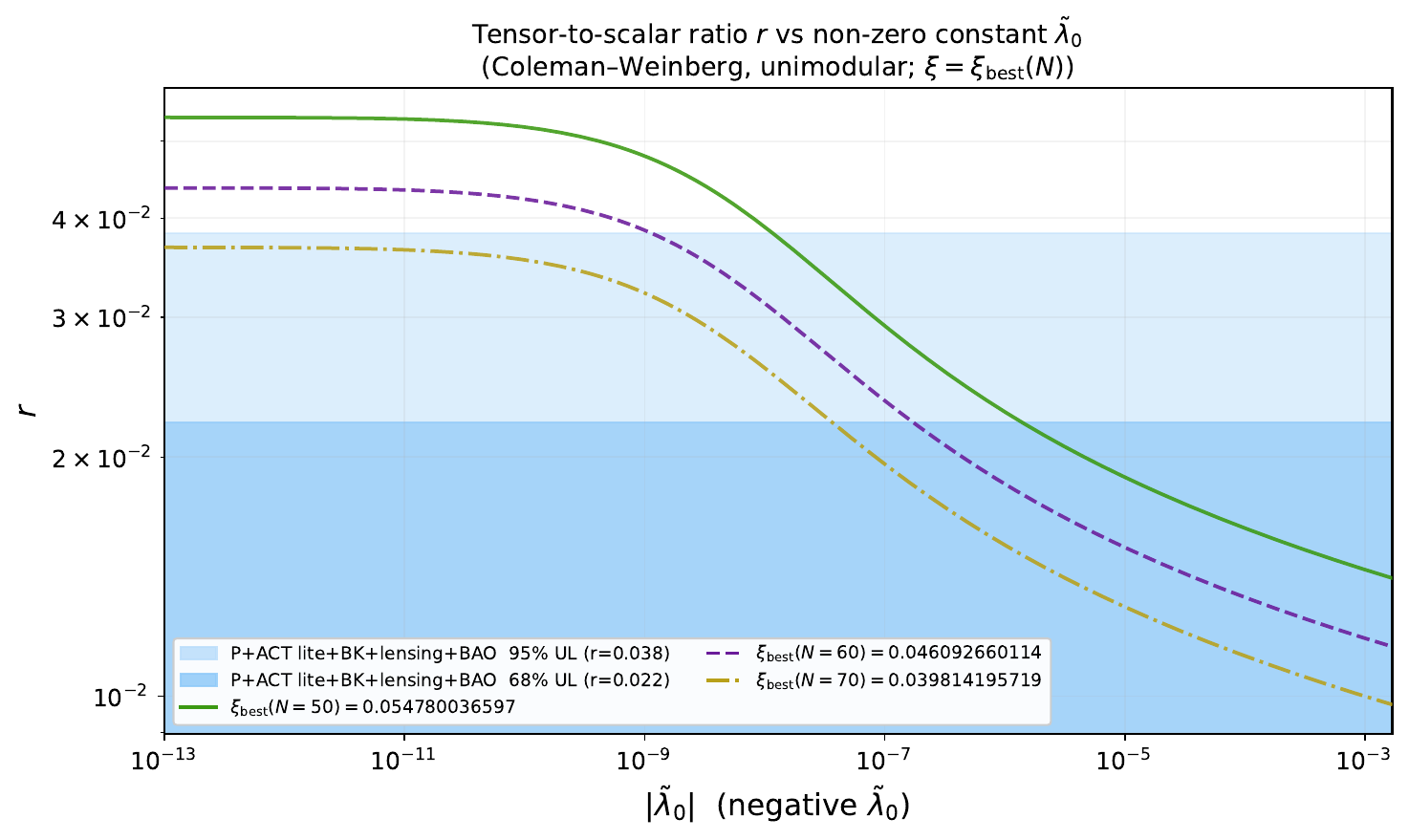}
\caption{(b) Tensor-to-scalar ratio $r$ as a function of the negative unimodular integration constant for the same large-field solutions as in \cref{fig:lambda_ns}. The trajectories show the correlated reduction of the tensor amplitude along the amplitude-normalized negative branch.}
\label{fig:lambda_r}
\end{figure}

\FloatBarrier
\begin{figure}[tbp]
\centering
\includegraphics[width=0.92\textwidth]{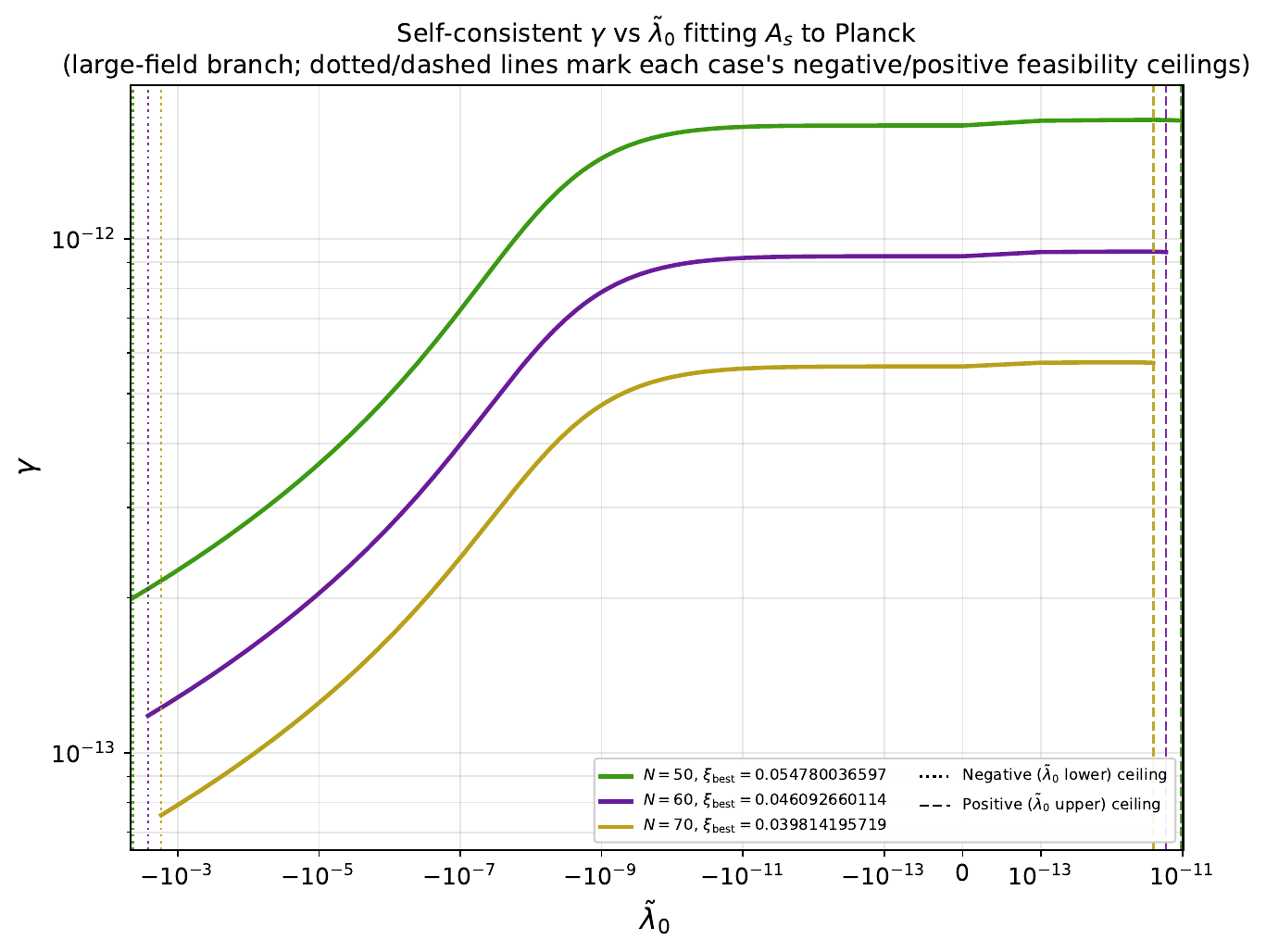}
\caption{The amplitude-normalized CW coupling $\gamma$ as a function of the unimodular integration constant $\lambdat$ for $N=50$, $60$, and $70$, each evaluated at its own preferred $\xi$. The positive and negative feasibility ceilings are case-dependent and are determined by the existence of a self-consistent solution satisfying the prescribed e-fold number and scalar amplitude.}
\label{fig:gamma_lambda}
\end{figure}

\FloatBarrier

\subsection{Observational posterior samples}
\label{subsec:obs_inputs}

The theoretical predictions are compared with three weighted posterior sample sets supplied with the numerical results: ACT lite+BK+lensing+BAO, Planck+BK+lensing+BAO, and Planck+ACT lite+BK+lensing+BAO. The data combinations are consistent with the observational programs represented by the Planck, BICEP/Keck, and ACT measurements \citep{Planck2020,BK2021,Aiola2020}. The weighted summaries used in the figures and the table are:
\begin{itemize}
\item ACT lite+BK+lensing+BAO: $\ns=0.9749$, with 68\% interval $[0.9682,0.9817]$ and 95\% interval $[0.9619,0.9883]$; the one-sided $r$ upper quantiles are $0.0219$ and $0.0379$ at 68\% and 95\%, respectively.
\item Planck+BK+lensing+BAO: $\ns=0.9699$, with 68\% interval $[0.9664,0.9736]$ and 95\% interval $[0.9630,0.9772]$; the corresponding $r$ upper quantiles are $0.0210$ and $0.0354$.
\item Planck+ACT lite+BK+lensing+BAO: $\ns=0.9741$, with 68\% interval $[0.9709,0.9772]$ and 95\% interval $[0.9676,0.9805]$; the corresponding $r$ upper quantiles are $0.0221$ and $0.0383$.
\end{itemize}
The two-dimensional posterior contours are obtained from the weighted samples with the GetDist analysis framework \citep{LewisBridle2002}. In the comparison below, the theoretical curves and swept regions represent predictions of the inflationary model, whereas the shaded contours represent posterior probability regions from the observational samples. The two types of regions have different meanings and are interpreted accordingly.

\section{Large-field numerical results}
\label{sec:large}

\subsection{Zero-multiplier baseline}

The $\lambdat=0$ solutions establish the reference predictions before the integration constant is varied. With $\xi$ selected according to \cref{eq:xi_best_values}, the amplitude-normalized values are
\begin{center}
\begin{tabular}{c c c c c c}
\toprule
$N$ & $\xi_{\rm best}$ & $\ns$ & $r$ & $\gamma$ & $(\phi_*,\phi_{\rm end})$\\
\midrule
50 & 0.05478 & 0.970845 & 0.053562 & $1.6613\times10^{-12}$ & $(29.3586,3.0296)$\\
60 & 0.04609 & 0.975883 & 0.043659 & $9.2476\times10^{-13}$ & $(32.7735,3.0903)$\\
70 & 0.03981 & 0.979449 & 0.036746 & $5.6440\times10^{-13}$ & $(35.8710,3.1371)$\\
\bottomrule
\end{tabular}
\end{center}
The tensor-to-scalar ratio decreases as the number of e-folds increases, while the scalar tilt moves closer to scale invariance. The baseline tensor amplitude nevertheless remains appreciable. In particular, the zero-multiplier solution does not by itself provide the same improvement as the negative integration-constant trajectories discussed below.

\subsection{Response to the unimodular integration constant}
\label{subsec:lambda_response}

For nonzero $\lambdat$, the potential changes through
\begin{equation}
U_{\eff}(\phi)=\frac{V_{\CW}(\phi)+\lambdat}{[1+\xi(\phi^2-\phi_0^2)]^2}.
\end{equation}
The numerator receives an additive contribution, while the denominator retains the nonminimal-coupling dependence. A negative integration constant reduces the potential energy over the relevant part of the trajectory and changes the logarithmic CW slope relative to the total potential. The net effect in the self-consistent solutions is a correlated increase in $\ns$ and decrease in $r$ over the negative branch. Both observables are recomputed after the normalization condition is imposed; the plotted trajectories are not obtained by shifting the zero-multiplier predictions by hand.

Figure~3(a) presents $\ns$ as a function of the negative integration constant, while Figure~3(b) presents the corresponding response of $r$. Displaying the two observables separately makes their dependence on the multiplier directly readable. For every point on these curves, $\gamma$, $\phi_*$, and $\phi_{\rm end}$ are readjusted to satisfy both $N$ and $\As$. The observational bands and upper-limit lines are defined by the weighted posterior summaries in \cref{subsec:obs_inputs}.

The combined behavior of $\ns$ and $r$ is the key reason for retaining the multiplier. The parameter does not merely shift the vertical normalization of the potential: because it is added before division by $F^2$, it changes the relative contribution of the CW term and hence the field derivatives that enter $\epsilon_V$ and $\eta_V$. The allowed range is finite, and the trend is not extrapolated beyond the self-consistent feasibility interval.

\subsection{Comparison in the \texorpdfstring{$(\ns,r)$}{(ns,r)} plane}
\label{subsec:nsr_comparison}

\begin{figure}[tbp]
\centering
\includegraphics[width=0.98\textwidth]{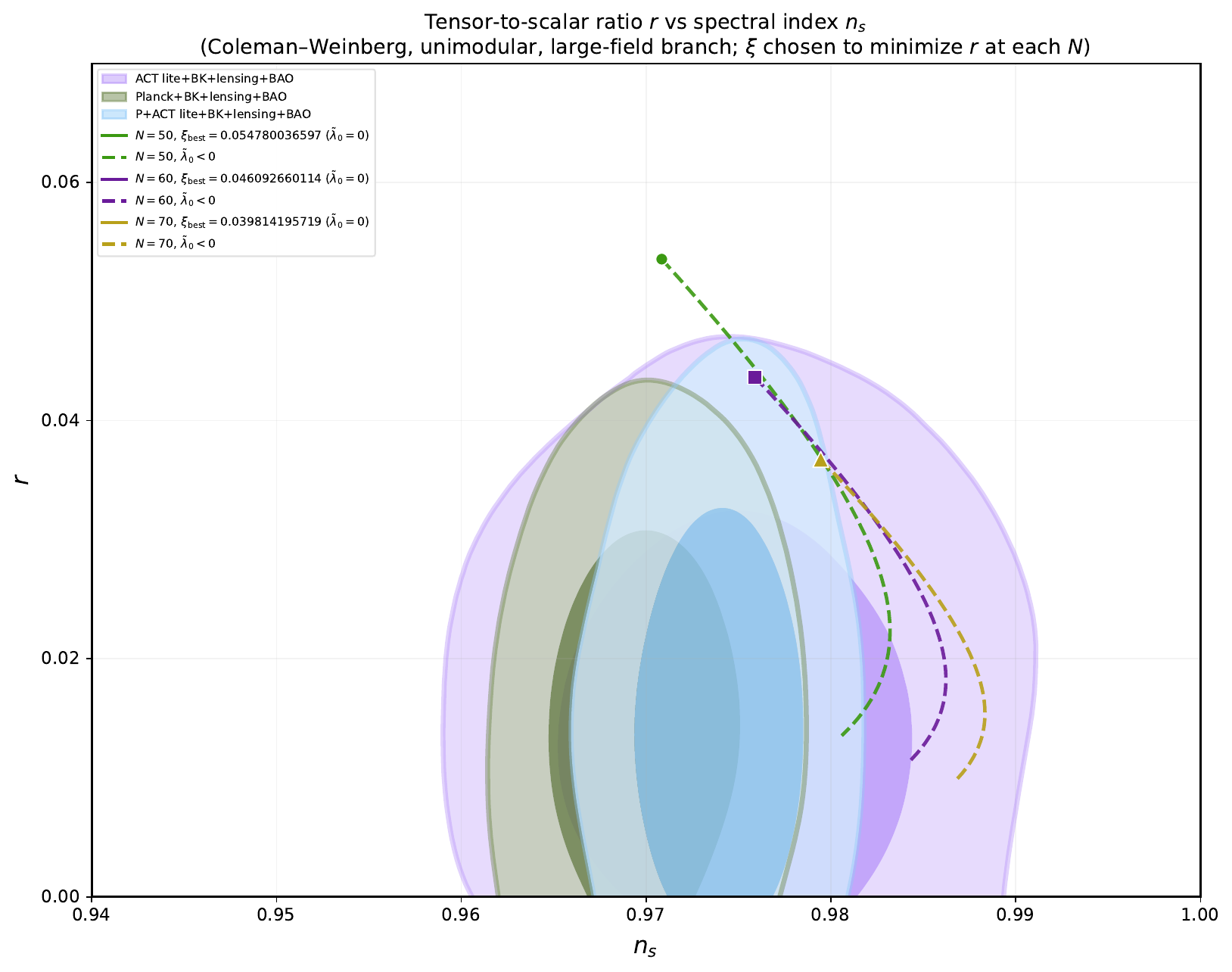}
\caption{Large-field theoretical predictions compared with the weighted posterior contours for ACT lite+BK+lensing+BAO, Planck+BK+lensing+BAO, and the combined Planck+ACT lite+BK+lensing+BAO samples. The baseline points and multiplier trajectories are evaluated at the preferred $\xi$ for each e-fold number.}
\label{fig:posterior_contours}
\end{figure}

A one-dimensional improvement in $\ns$ or $r$ does not by itself establish observational viability. The appropriate comparison is made in the two-dimensional plane, where the theoretical trajectories can be located relative to the weighted posterior contours. \Cref{fig:posterior_contours} displays the three posterior contour sets together with the model predictions. It shows how the negative-$\lambdat$ trajectories leave the zero-multiplier baseline and move toward the observationally preferred region. The amount of overlap depends on the e-fold number and on the data combination, so the multiplier should be interpreted as an additional dynamical handle rather than as a parameter that makes every baseline prediction viable.

\FloatBarrier

\begin{figure}[tbp]
\centering
\includegraphics[width=0.98\textwidth]{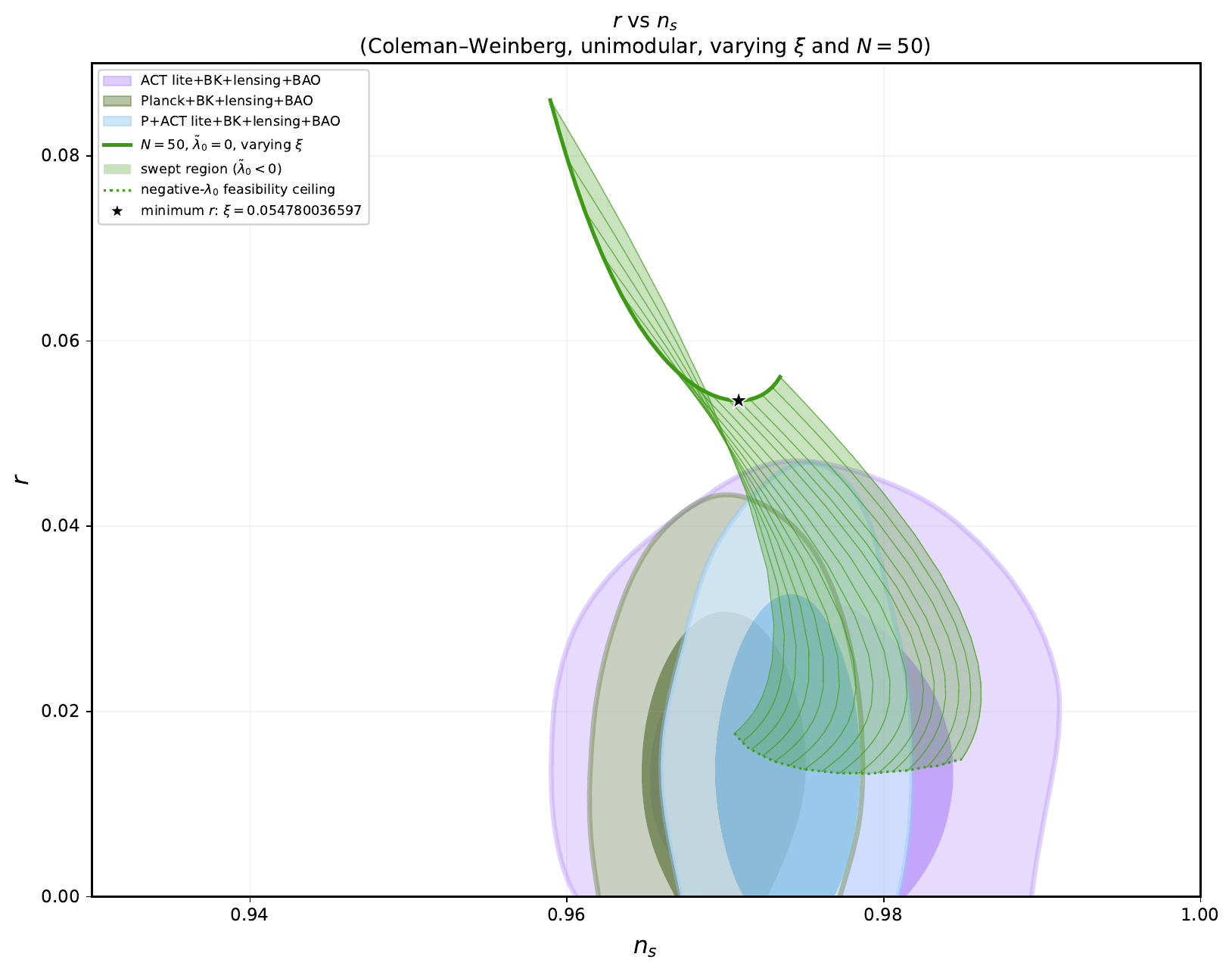}
\caption{The $N=50$ filled theoretical region in the $(\ns,r)$ plane obtained by varying $\xi$ and following the negative-$\lambdat$ solutions to their individual feasibility ceilings. The green zero-multiplier trajectory is the baseline boundary, the shaded sweep is generated by the allowed negative multiplier, and the star marks the minimum-$r$ point at $\xi_{\rm best}=0.05478$. The overlap with the supplied posterior contours displays the central result: the unimodular integration constant opens an observationally relevant region in the large-field CW model.}
\label{fig:lambda_surface}
\end{figure}

The central result of the paper is visualized by the filled $N=50$ region in \cref{fig:lambda_surface}. Here $N$ is held fixed while $\xi$ is scanned, and for each value of $\xi$ the negative-multiplier trajectory is followed to its own feasibility boundary. The zero-multiplier curve forms one boundary of the family, while the solutions at the negative feasibility ceilings form the other. The fill is the union of model trajectories generated as both $\xi$ and $\lambdat$ vary within the self-consistent domain.

\FloatBarrier
\FloatBarrier

This figure condenses the principal physical mechanism: the Einstein-frame minimally coupled scalar sector and the unimodular integration constant act together. The zero-multiplier curve alone traces the predictions of the CW model with the selected scalar--curvature coupling in the original formulation, while the additional multiplier direction sweeps a finite region in the $(\ns,r)$ plane. The observational contours overlap part of this swept region. Consequently, the unimodular integration constant can open a path into the data-preferred region even when the corresponding zero-multiplier prediction is displaced from it. The filled surface is a set of theoretical predictions, not a posterior confidence region; its meaning is dynamical and parameter-geometric, not probabilistic.

The construction can be read as a two-parameter map. For every sampled value of $\xi$ in the prescribed interval, the calculation first solves the zero-multiplier model and then moves along the negative-$\lambdat$ branch. The negative endpoint is not shared between different values of $\xi$: it is found from the last amplitude-normalized solution on that particular trajectory. Each pair $(\xi,\lambdat)$ is therefore mapped to a point $(\ns(\xi,\lambdat),r(\xi,\lambdat))$. Connecting neighboring points in the scan produces the shaded image of the allowed parameter domain. The boundary at $\lambdat=0$ is the family of ordinary nonzero-$\xi$ CW predictions, whereas the opposite boundary is the family of individual negative-multiplier ceilings. The two side boundaries arise from the endpoints of the $\xi$ scan.

This representation is more informative than plotting a single optimized-$\xi$ curve. A single curve can show that a particular multiplier value improves one selected trajectory, but it cannot show whether that behavior persists when the scalar--curvature coupling parameter is changed. The filled region tests precisely this question: it shows the combined leverage of the two model parameters while respecting the normalization and branch conditions at every point. The star in \cref{fig:lambda_surface} at $\xi=0.05478$ identifies the zero-multiplier minimum-$r$ baseline within this family; the shaded area around it displays the additional predictions generated by the unimodular degree of freedom. In this sense, the surface plot is the central visual result of the paper, because it demonstrates how retaining the integration constant changes the \emph{set} of accessible inflationary predictions rather than merely shifting one reference point.

The comparison with the posterior contours should be interpreted geometrically. An overlap means that at least part of the theoretical image lies in a region favored by the corresponding weighted data distribution. It does not mean that every point in the filled region is equally likely, nor that the model parameters have been marginalized into a Bayesian posterior by the plot itself. The contour plot supplies the observational probability information; the filled surface supplies the map of theoretical predictions. Together they answer the main question of the work: whether the unimodular integration constant can enlarge the region of the large-field CW model that is compatible with the observed scalar and tensor perturbations.

\subsection{Allowed parameter intervals: interpretation of Table 1}
\label{subsec:allowed_tables}

Table \ref{tab:ranges} and gives the parameter intervals associated with the 68\% ($1\sigma$) and 95\% ($2\sigma$) marginal scalar-index ranges. It is obtained by mapping each observational $\ns$ interval onto the amplitude-normalized model trajectory at fixed $N$ and $\xi_{\rm best}(N)$. For a selected data set and confidence level, define the accepted multiplier subset by
\begin{equation}
\mathcal I_{\lambda}^{(p)}=
\left\{\lambdat\in\mathcal I_{\rm feasible}:
\ns(\lambdat)\in\mathcal I_{\ns}^{(p)}\right\},\qquad p=68\%,95\%.
\label{eq:lambda_interval_set}
\end{equation}
The first interval column in each part is the image of this subset in $\lambdat$. The corresponding ranges of $r$, $\phi_*$, and $\phi_{\rm end}$ are then evaluated over the same model points. The field values are quoted in the original scalar coordinate $\phi$, for which the coupling and CW potential have their simplest analytic forms.

This construction is a direct projection of the numerical solution set. At fixed $N$ and $\xi_{\rm best}$, each allowed value of $\lambdat$ labels a fully normalized model point, and that point carries a definite quadruple $(\ns,r,\phi_*,\phi_{\rm end})$. The observational scalar-tilt interval selects a connected segment, or no segment of this family. The table then reports the minimum and maximum of each remaining parameter over precisely that selected segment. Thus the entries in a row are correlated ranges generated by the same set of solutions; they are not independent parameter priors and should not be combined by selecting arbitrary endpoints from different columns.

The 95\% interval is wider and can therefore admit solutions that are absent from the 68\% selection. This is visible, for example, for $N=60$ with the Planck+BK+lensing+BAO combination: the 68\% scalar interval has no intersection with the model trajectory, while the 95\% interval admits a finite interval. For $N=70$, the Planck-only row has no intersection at either level. These are statements about the position of the theoretical curve relative to the measured marginal scalar tilt, not failures of the inflationary background solution itself.

The large excursions in $\phi_*$ and $\phi_{\rm end}$ in some ACT rows have a dynamical origin. As the multiplier approaches a feasibility ceiling, the normalized potential becomes increasingly close to the edge of the branch on which the prescribed $N$ and scalar amplitude can be satisfied. The horizon-crossing and end points can then move far from their baseline values. This is why the field ranges are reported alongside $\lambdat$ and $r$: they make visible the change in the inflationary trajectory that accompanies the improvement in the scalar tilt. The table therefore contains more information than a list of acceptable multiplier values; it records how observational selection maps into the underlying field evolution.

The table is organized by e-fold number. Each $N$ has its own optimized coupling, so the same numerical value of $\xi$ is used for the three data combinations within that block. The data combinations are kept separate because their marginal $\ns$ intervals differ. A row labelled ``Outside bounds'' means that the corresponding theoretical trajectory does not intersect the specified one-dimensional marginal $\ns$ interval anywhere in its feasible multiplier range. It does not denote an independent prior on $\lambdat$; it records the absence of an intersection between a model curve and an observational interval.

The table must be read together with the two-dimensional comparison in \cref{subsec:nsr_comparison}. They are constructed from the one-dimensional $\ns$ marginal intervals and then report the induced ranges of $r$ and field values. Therefore, a nonempty entry is a necessary scalar-tilt match, not by itself a sufficient joint $(\ns,r)$ match. This is why the posterior-contour figure and the $N=50$ swept region are essential complements to the table. In rows where the $r$ interval extends above the corresponding one-sided observational upper quantile, only the portion also lying inside the joint posterior region is relevant for a simultaneous comparison.

The distinction between the multiplier feasibility interval and the table intervals is also important. The feasibility interval is determined by the dynamical and amplitude-normalization equations before applying observational constraints. The table interval is the subset of that feasible interval whose predicted scalar tilt falls inside a chosen observational marginal range. Thus the feasibility interval answers whether a normalized inflationary solution exists, whereas the tabulated interval answers where that solution also satisfies a specified scalar-tilt condition.

\clearpage
\begin{sidewaystable}[p]
\centering
\caption{Allowed large-field parameter intervals selected by the 68\% ($1\sigma$) and 95\% ($2\sigma$) marginal scalar-index ranges.}
\label{tab:ranges}
\scriptsize
\setlength{\tabcolsep}{3pt}
\renewcommand{\arraystretch}{1.15}
\resizebox{\linewidth}{!}{%
\begin{tabular}{c c c @{\hspace{12pt}} c c c c @{\hspace{12pt}} c c c c}
\toprule
 & & & \multicolumn{4}{c}{68\% ($1\sigma$)} & \multicolumn{4}{c}{95\% ($2\sigma$)}\\
\cmidrule(lr){4-7}\cmidrule(lr){8-11}
$\xi_{\rm best}$ & $N$ & Data combination
 & $\widetilde\lambda_0$ range & $r$ range & $\phi_*$ range & $\phi_{\rm end}$ range
 & $\widetilde\lambda_0$ range & $r$ range & $\phi_*$ range & $\phi_{\rm end}$ range\\
\midrule
\multirow{3}{*}{0.05478} & \multirow{3}{*}{50}
 & \shortstack{ACT lite+BK+\\lensing+BAO}
 & $[-4.27\times10^{-3},0]$ & $[0.0135,0.0536]$ & $[29.36,747.24]$ & $[3.03,282.86]$
 & $[-4.27\times10^{-3},0]$ & $[0.0135,0.0536]$ & $[29.36,747.24]$ & $[3.03,282.86]$\\[10pt]
 & & \shortstack{Planck+BK+\\lensing+BAO}
 & $[-6.10\times10^{-10},0]$ & $[0.0492,0.0536]$ & $[29.36,31.43]$ & $[3.03,5.25]$
 & $[-4.69\times10^{-9},0]$ & $[0.0423,0.0536]$ & $[29.36,36.04]$ & $[3.03,8.03]$\\[10pt]
 & & \shortstack{Planck+ACT+BK+\\lensing+BAO}
 & $[-4.69\times10^{-9},-3.39\times10^{-12}]$ & $[0.0423,0.0535]$ & $[29.40,36.04]$ & $[3.10,8.03]$
 & $[-2.49\times10^{-8},0]$ & $[0.0349,0.0536]$ & $[29.36,44.86]$ & $[3.03,12.08]$\\[10pt]
\midrule
\multirow{3}{*}{0.04609} & \multirow{3}{*}{60}
 & \shortstack{ACT lite+BK+\\lensing+BAO}
 & $[-5.65\times10^{-9},0]$ & $[0.0333,0.0437]$ & $[32.77,41.87]$ & $[3.09,9.48]$
 & $[-2.45\times10^{-3},0]$ & $[0.0114,0.0437]$ & $[32.77,757.12]$ & $[3.09,278.90]$\\[10pt]
 & & \shortstack{Planck+BK+\\lensing+BAO}
 & \multicolumn{4}{c}{Outside bounds}
 & $[-2.00\times10^{-10},0]$ & $[0.0415,0.0437]$ & $[32.77,34.10]$ & $[3.09,4.77]$\\[10pt]
 & & \shortstack{Planck+ACT+BK+\\lensing+BAO}
 & $[-2.00\times10^{-10},0]$ & $[0.0415,0.0437]$ & $[32.77,34.10]$ & $[3.09,4.77]$
 & $[-2.69\times10^{-9},0]$ & $[0.0358,0.0437]$ & $[32.77,38.89]$ & $[3.09,7.97]$\\[10pt]
\midrule
\multirow{3}{*}{0.03981} & \multirow{3}{*}{70}
 & \shortstack{ACT lite+BK+\\lensing+BAO}
 & $[-6.10\times10^{-10},0]$ & $[0.0332,0.0367]$ & $[35.87,39.05]$ & $[3.14,6.40]$
 & $[-1.69\times10^{-3},0]$ & $[0.00976,0.0367]$ & $[35.87,783.76]$ & $[3.14,282.80]$\\[10pt]
 & & \shortstack{Planck+BK+\\lensing+BAO}
 & \multicolumn{4}{c}{Outside bounds}
 & \multicolumn{4}{c}{Outside bounds}\\[10pt]
 & & \shortstack{Planck+ACT+BK+\\lensing+BAO}
 & \multicolumn{4}{c}{Outside bounds}
 & $[-1.38\times10^{-10},0]$ & $[0.0351,0.0367]$ & $[35.87,37.17]$ & $[3.14,4.87]$\\[10pt]
\bottomrule
\end{tabular}%
}
\end{sidewaystable}
\clearpage

\section{Small-field branch: a scalar-tilt limitation}
\label{sec:small}

The same CW potential supports inflation on the small-field side of the radiative minimum, $0<\phi<\phi_0$. This branch has an observational pattern opposite to the large-field case. The tensor-to-scalar ratio is already extremely small, with the numerical trajectories remaining of order $10^{-7}$, so an excessive tensor signal is not the limiting factor. The challenge is the scalar spectral index: for the small-field couplings used in the numerical scan, the baseline predictions lie in the range $0.940\leq\ns\leq0.957$, below the combined Planck+ACT lite+BK+lensing+BAO 68\% interval $[0.9709,0.9772]$ and its 95\% interval $[0.9676,0.9805]$.

The negative or positive multiplier deformation in the small-field scan moves $\ns$ in the direction of the observed interval, but the scalar-amplitude normalization restricts the multiplier to a very narrow range. Across the amplitude-compatible scan the change in $\ns$ is small, and the curves remain below the lower edge of the 95\% combined-data interval. Reaching the observational band would require a larger multiplier contribution than is compatible with the simultaneous solution of the e-fold and scalar-amplitude conditions. Hence the small-field branch is not limited by $r$; it is limited by a scalar tilt that remains too red within the amplitude-normalized multiplier domain.

\begin{figure}[H]
\centering
\includegraphics[width=0.86\textwidth]{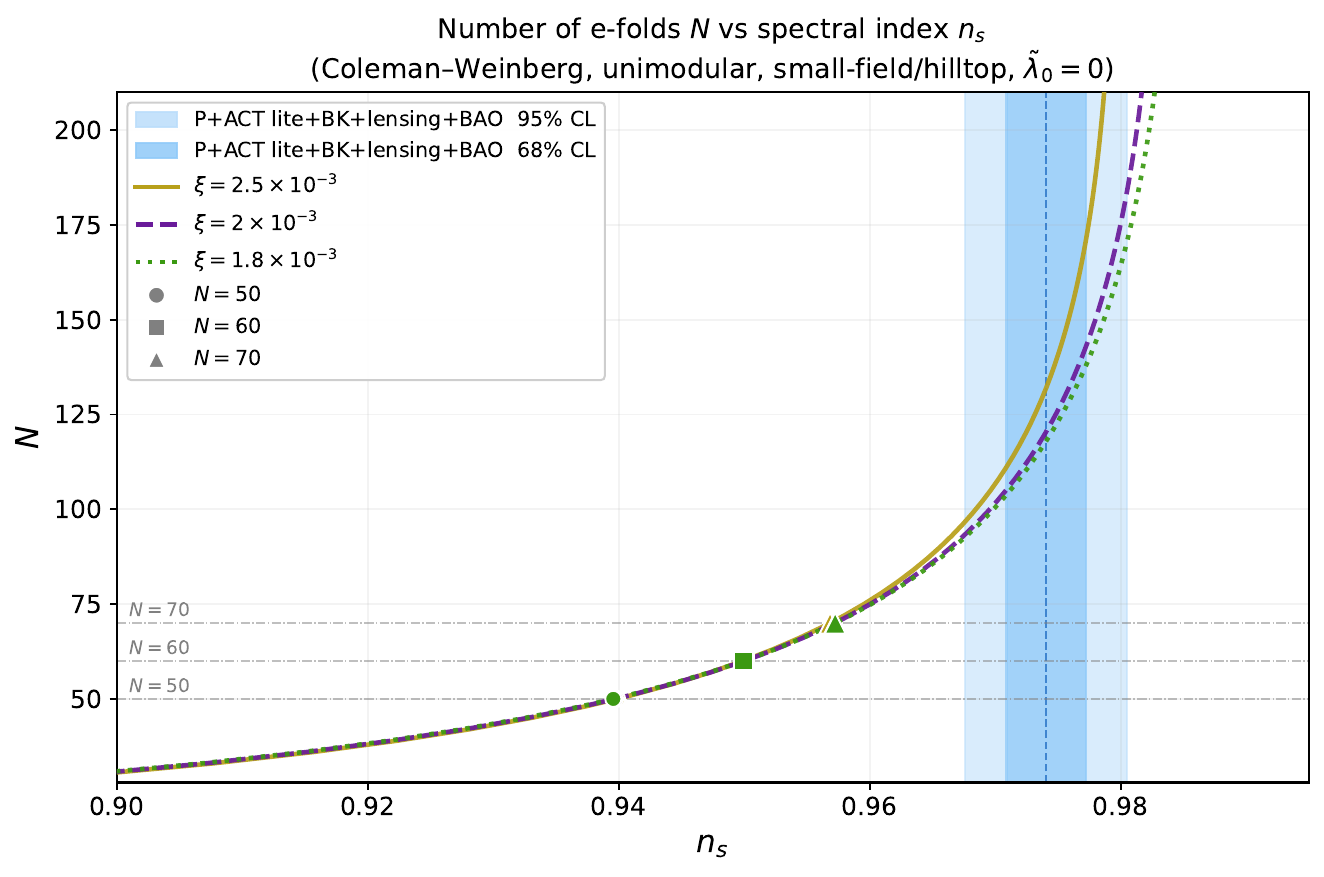}
\caption{Small-field baseline relation between the e-fold number $N$ and the scalar spectral index at $\lambdat=0$. The selected small-field values of $\xi$ give a scalar tilt below the 68\% and 95\% bands of the combined posterior samples.}
\label{fig:small_N_ns}
\end{figure}

\begin{figure}[tbp]
\centering
\includegraphics[width=0.90\textwidth]{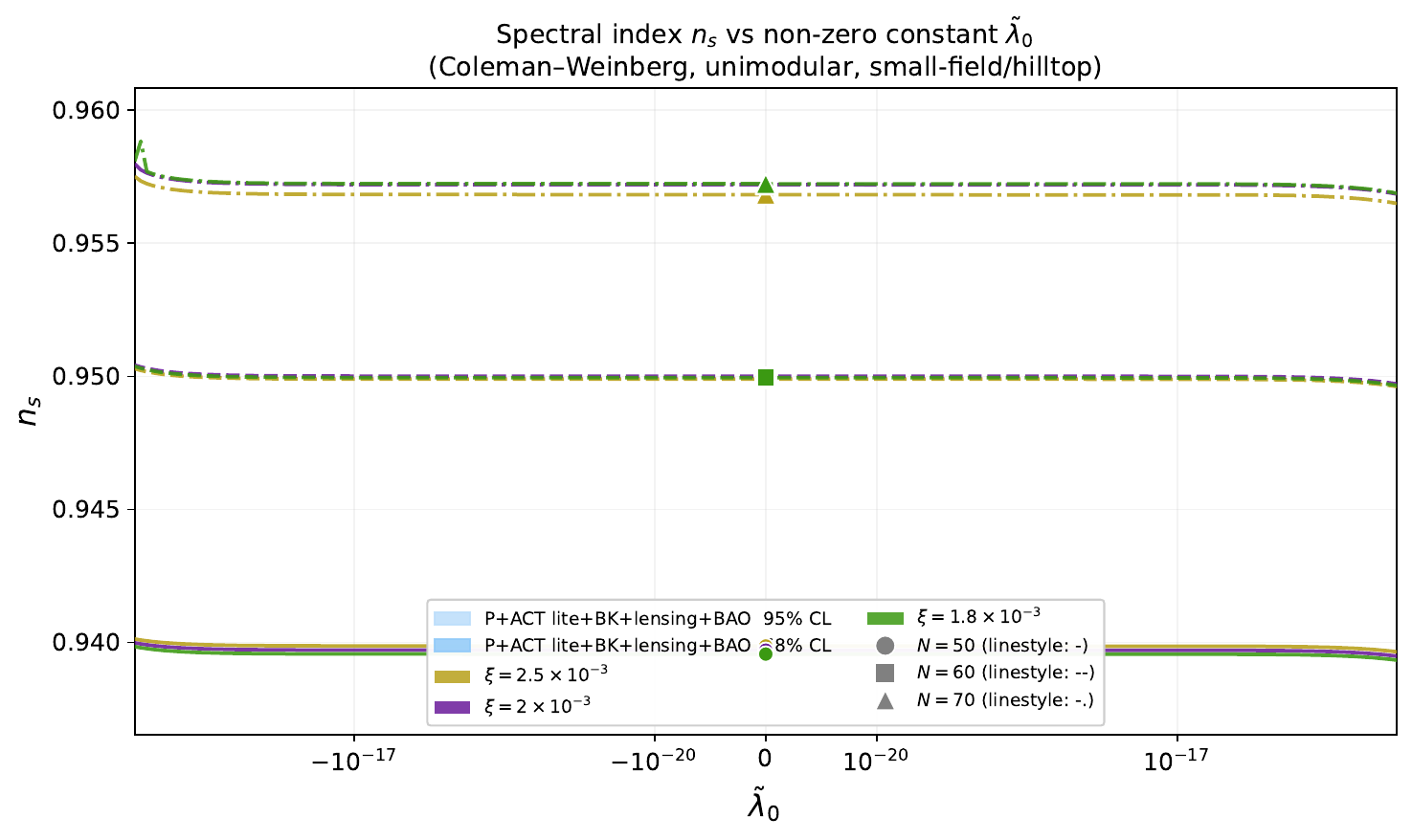}
\caption{Small-field scalar spectral index as a function of the unimodular integration constant. The plotted curves show the amplitude-normalized solutions for the selected small-field couplings. The multiplier shifts $\ns$ toward larger values, but the allowed normalized trajectories remain below the combined-data scalar-tilt interval.}
\label{fig:small_ns_lambda}
\end{figure}

\FloatBarrier
\section{Conclusions}
\label{sec:conclusions}

We have studied Coleman--Weinberg inflation in unimodular gravity with the vacuum-shifted scalar--curvature coupling defined in \cref{eq:F_shifted}, using an Einstein-frame formulation for the inflationary analysis. The conformal transformation puts the gravitational sector in the standard Einstein--Hilbert form, so the scalar is minimally coupled to gravity in this frame. Its kinetic term is made canonical by the subsequent scalar-field redefinition described in \cref{subsec:scalar_field_redefinition}.

The unimodular condition is imposed with an Lagrange multiplier. Its on-shell constancy follows from the Einstein-frame scalar equation and the contracted Bianchi identity, as shown in \cref{eq:lambda_constant_E}. The resulting integration constant contributes to the effective potential given in \cref{eq:Ueff}. This contribution is retained rather than set to zero, allowing its effect on the inflationary observables to be assessed.

On the large-field branch, the preferred values of the coupling parameter are selected separately for $N=50$, $60$, and $70$ by minimizing the tensor-to-scalar ratio on the zero-multiplier slice. The resulting values are $0.05478$, $0.04609$, and $0.03981$. At these baselines the tensor-to-scalar ratio decreases with increasing $N$, but remains appreciable. For nonzero integration constant, the scalar amplitude is imposed at every scan point, so the CW normalization, horizon-crossing field, and end-of-inflation field are recalculated self-consistently. The resulting feasibility interval is strongly asymmetric: the negative branch spans a much larger absolute range than the positive branch.

The negative-multiplier trajectories move toward larger scalar spectral index and smaller tensor-to-scalar ratio. The posterior-contour comparison and the filled $N=50$ theoretical region in the $(n_s,r)$ plane show that varying the coupling parameter together with the unimodular integration constant produces a finite region that overlaps the observational contours. The table provides the complementary mapping between the marginal scalar-index intervals and the corresponding multiplier, tensor-to-scalar-ratio, and field-value intervals.

The small-field branch has a different limitation. Its tensor amplitude is already negligible, but its scalar spectral index remains below the observational band. Although the multiplier shifts the tilt in the favorable direction, the amplitude-normalized domain does not provide a sufficiently large shift. Thus, within the parameter ranges and approximations studied here, the unimodular integration constant improves the observational reach of the large-field branch, whereas the small-field branch remains limited by its scalar tilt.

\section{Future outlook}
Several extensions can test the robustness and physical origin of these results. First, the analysis can be extended beyond the leading-order potential slow-roll approximation by solving the background equations and the Mukhanov--Sasaki mode equation numerically for representative points across the allowed parameter domain. This would establish the size of higher-order corrections to $(n_s,r)$ and test whether the overlap with the observational contours persists. Second, the scalar--tensor model can be studied with alternative radiatively generated potentials and with renormalization-group-improved Coleman--Weinberg potentials, to determine whether the improvement is specific to the logarithmic potential used here or is shared by a wider class of radiative models. Finally, a systematic investigation of reheating and the resulting uncertainty in the number of e-folds could connect the predicted observables to a self-consistent post-inflationary history.

\section{Declaration of competing interest}
The authors declare that they have no known competing financial interests or personal relationships that could have appeared to influence the work reported in this paper.
\section{Funding declaration}
The authors declare that there was no funding.
\section{Data availability}
\url{https://lambda.gsfc.nasa.gov/product/}, See also references section.

\clearpage
\bibliographystyle{unsrtnat}
\bibliography{references}

@article{Einstein1919,
  author = {Einstein, Albert},
  title = {Spielen Gravitationsfelder im Aufbau der materiellen Elementarteilchen eine wesentliche Rolle?},
  journal = {Sitzungsberichte der Preussischen Akademie der Wissenschaften, Physikalisch-Mathematische Klasse},
  pages = {349--356}, year = {(1919)}
}

@book{PauliRelativity,
  author = {Pauli, Wolfgang}, title = {Theory of Relativity}, publisher = {Dover}, address = {New York}, year = {(1981)}, note = {English reprint}
}

@article{AndersonFinkelstein1971,
  author = {Anderson, John L. and Finkelstein, David},
  title = {Cosmological Constant and Fundamental Length},
  journal = {American Journal of Physics}, volume = {39}, pages = {901--904}, year = {(1971)},
  note = {Bibliographic details as listed in Alvarez and Velasco-Aja (2023)}
}

@article{VanDerBij1982,
  author = {van der Bij, J. J. and van Dam, H. and Ng, Y. J.},
  title = {The exchange of massless spin-two particles},
  journal = {Physica A}, volume = {116}, pages = {307--320}, year = {(1982)}
}

@article{BuchmullerDragon1988,
  author = {Buchmuller, W. and Dragon, N.},
  title = {Einstein gravity from restricted coordinate invariance},
  journal = {Physics Letters B}, volume = {207}, pages = {292--294}, year = {(1988)}
}

@article{BuchmullerDragon1989,
  author = {Buchmuller, W. and Dragon, N.},
  title = {Gauge fixing and the cosmological constant},
  journal = {Physics Letters B}, volume = {223}, pages = {313--317}, year = {(1989)}
}

@article{HenneauxTeitelboim1989,
  author = {Henneaux, Marc and Teitelboim, Claudio},
  title = {The cosmological constant and general covariance},
  journal = {Physics Letters B}, volume = {222}, number = {2}, pages = {195--199}, year = {(1989)},
  doi = {10.1016/0370-2693(89)91251-3}
}

@article{Unruh1989,
  author = {Unruh, William G.},
  title = {Unimodular theory of canonical quantum gravity},
  journal = {Physical Review D}, volume = {40}, pages = {1048--1052}, year = {(1989)},
  doi = {10.1103/PhysRevD.40.1048}
}

@article{Weinberg1989,
  author = {Weinberg, Steven},
  title = {The cosmological constant problem},
  journal = {Reviews of Modern Physics}, volume = {61}, pages = {1--23}, year = {(1989)},
  doi = {10.1103/RevModPhys.61.1}
}

@article{Carroll2001,
  author = {Carroll, Sean M.},
  title = {The cosmological constant},
  journal = {Living Reviews in Relativity}, volume = {4}, number = {1}, year = {(2001)},
  doi = {10.12942/lrr-2001-1}
}

@article{UnruhWald1989,
  author = {Unruh, William G. and Wald, Robert M.},
  title = {Time and the interpretation of canonical quantum gravity},
  journal = {Physical Review D}, volume = {40}, pages = {2598--2614}, year = {(1989)}
}

@article{Alvarez2005,
  author = {Alvarez, Enrique},
  title = {Can one tell Einstein's unimodular theory from Einstein's general relativity?},
  journal = {Journal of High Energy Physics}, volume = {2005}, number = {03}, pages = {002}, year = {(2005)},
  doi = {10.1088/1126-6708/2005/03/002}, eprint = {hep-th/0501146}
}

@article{AlvarezBlasGarrigaVerdaguer2006,
  author = {Alvarez, Enrique and Blas, Diego and Garriga, Jaume and Verdaguer, Enric},
  title = {Transverse Fierz--Pauli symmetry},
  journal = {Nuclear Physics B}, volume = {756}, pages = {148--170}, year = {(2006)},
  doi = {10.1016/j.nuclphysb.2006.09.004}, eprint = {hep-th/0606019}
}

@article{AlvarezVidal2010,
  author = {Alvarez, Enrique and Vidal, Roberto},
  title = {Weyl transverse gravity and the cosmological constant},
  journal = {Physical Review D}, volume = {81}, pages = {084057}, year = {(2010)},
  doi = {10.1103/PhysRevD.81.084057}
}

@article{EllisElstMuruganUzan2011,
  author = {Ellis, George F. R. and Elst, Henk van and Murugan, Jeff and Uzan, Jean-Philippe},
  title = {On the trace-free Einstein equations as a viable alternative to general relativity},
  journal = {Classical and Quantum Gravity}, volume = {28}, pages = {225007}, year = {(2011)}
}

@article{AlvarezHerreroValea2013,
  author = {Alvarez, Enrique and Herrero-Valea, Mario},
  title = {Unimodular gravity with external sources},
  journal = {Journal of Cosmology and Astroparticle Physics}, volume = {2013}, number = {01}, pages = {014}, year = {(2013)},
  doi = {10.1088/1475-7516/2013/01/014}, eprint = {1209.6223}
}

@article{SaezGomez2016,
  author = {Saez-Gomez, Diego},
  title = {Analyzing modified unimodular gravity via Lagrange multipliers},
  journal = {Physical Review D}, volume = {93}, pages = {124040}, year = {(2016)},
  doi = {10.1103/PhysRevD.93.124040}, eprint = {1602.04771}
}

@article{NojiriOdintsovOikonomou2016,
  author = {Nojiri, Shin'ichi and Odintsov, Sergei D. and Oikonomou, V. K.},
  title = {Unimodular {f(R)} gravity},
  journal = {Journal of Cosmology and Astroparticle Physics}, volume = {2016}, number = {05}, pages = {046}, year = {(2016)}
}

@article{AlvarezGonzalezMartinMartin2016,
  author = {Alvarez, Enrique and Gonzalez-Martin, Sergio and Martin, C. P.},
  title = {Note on the gauge symmetries of unimodular gravity},
  journal = {Physical Review D}, volume = {93}, pages = {123018}, year = {(2016)}
}

@article{CarballoRubio2022,
  author = {Carballo-Rubio, Raquel and Garay, Luis J. and Garcia-Moreno, Gonzalo},
  title = {Unimodular gravity vs general relativity: a status report},
  journal = {Classical and Quantum Gravity}, volume = {39}, pages = {243001}, year = {(2022)}
}

@article{Jirousek2023,
  author = {Jirousek, Pavel},
  title = {Unimodular approaches to the cosmological constant problem},
  journal = {Universe}, volume = {9}, pages = {84}, year = {(2023)}
}

@article{AlvarezVelascoAja2023,
  author = {Alvarez, Enrique and Velasco-Aja, Eduardo},
  title = {A primer on unimodular gravity},
  year = {(2023)}, eprint = {2301.07641}, archivePrefix = {arXiv}, primaryClass = {gr-qc},
  note = {Invited chapter for the Handbook of Quantum Gravity}
}

@article{MalekpourNozariRajabiRashidi2024,
  author = {Malekpour, Manda and Nozari, Kourosh and Rajabi, Fateme and Rashidi, Narges},
  title = {Non-minimal unimodular inflation},
  journal = {Physics of the Dark Universe}, volume = {43}, pages = {101405}, year = {(2024)}
}

@article{NozariMalekpour2024,
  author = {Nozari, Kourosh and Malekpour, Manda},
  title = {Higgs inflation in unimodular gravity},
  journal = {Progress of Theoretical and Experimental Physics}, volume = {2024}, number = {6}, pages = {063E02}, year = {(2024)},
  doi = {10.1093/ptep/ptae085}
}

@article{Guth1981,
  author = {Guth, Alan H.},
  title = {The inflationary universe: A possible solution to the horizon and flatness problems},
  journal = {Physical Review D}, volume = {23}, pages = {347--356}, year = {(1981)},
  doi = {10.1103/PhysRevD.23.347}
}

@article{Linde1982,
  author = {Linde, Andrei D.},
  title = {A new inflationary universe scenario: A possible solution of the horizon, flatness, homogeneity, isotropy and primordial monopole problems},
  journal = {Physics Letters B}, volume = {108}, pages = {389--393}, year = {(1982)}
}

@article{AlbrechtSteinhardt1982,
  author = {Albrecht, Andreas and Steinhardt, Paul J.},
  title = {Cosmological consequences of a new inflationary universe},
  journal = {Physical Review Letters}, volume = {48}, pages = {1220--1223}, year = {(1982)}
}

@article{Starobinsky1980,
  author = {Starobinsky, Alexei A.},
  title = {A new type of isotropic cosmological models without singularity},
  journal = {Physics Letters B}, volume = {91}, pages = {99--102}, year = {(1980)}
}

@article{MukhanovChibisov1981,
  author = {Mukhanov, Viatcheslav F. and Chibisov, G. V.},
  title = {Quantum fluctuations and a nonsingular universe},
  journal = {JETP Letters}, volume = {33}, pages = {532--535}, year = {(1981)}
}

@article{ColemanWeinberg1973,
  author = {Coleman, Sidney R. and Weinberg, Erick J.},
  title = {Radiative corrections as the origin of spontaneous symmetry breaking},
  journal = {Physical Review D}, volume = {7}, pages = {1888--1910}, year = {(1973)},
  doi = {10.1103/PhysRevD.7.1888}
}

@article{ShafiVilenkin1984,
  author = {Shafi, Qaisar and Vilenkin, Alexander},
  title = {Inflation with {SU(5)}},
  journal = {Physical Review Letters}, volume = {52}, pages = {691--694}, year = {(1984)},
  doi = {10.1103/PhysRevLett.52.691}
}

@article{SenoguzShafi2006,
  author = {Senoguz, V. N. and Shafi, Q.},
  title = {Coleman--Weinberg potential in good agreement with {WMAP}},
  journal = {Physical Review D}, volume = {73}, pages = {127301}, year = {(2006)},
  doi = {10.1103/PhysRevD.73.127301}, eprint = {hep-ph/0603830}
}

@article{BarenboimChunLee2014,
  author = {Barenboim, Gabriela and Chun, Eung Jin and Lee, Hyun Min},
  title = {Coleman--Weinberg inflation in light of {Planck}},
  journal = {Physics Letters B}, volume = {730}, pages = {81--88}, year = {(2014)},
  doi = {10.1016/j.physletb.2014.01.039}, eprint = {1309.1695}
}

@article{RehmanShafiWickman2008,
  author = {Rehman, Mansoor Ur and Shafi, Qaisar and Wickman, James R.},
  title = {GUT inflation, radiative corrections and the {WMAP} three year data},
  journal = {Physical Review D}, volume = {78}, pages = {123516}, year = {(2008)}
}

@article{KannikeRacioppiRaidal2014,
  author = {Kannike, Kristjan and Racioppi, Antonio and Raidal, Martti},
  title = {Embedding inflation into the standard model: More evidence for classical scale invariance},
  journal = {Journal of High Energy Physics}, volume = {2014}, number = {06}, pages = {154}, year = {(2014)},
  doi = {10.1007/JHEP06(2014)154}, eprint = {1405.3987}
}

@article{BezrukovShaposhnikov2008,
  author = {Bezrukov, Fedor L. and Shaposhnikov, Mikhail},
  title = {The standard model higgs boson as the inflaton},
  journal = {Physics Letters B}, volume = {659}, pages = {703--706}, year = {(2008)},
  doi = {10.1016/j.physletb.2007.11.072}, eprint = {0710.3755}
}

@article{FakirUnruh1990,
  author = {Fakir, R. and Unruh, W. G.},
  title = {Improvement on cosmological inflation by nonminimal coupling},
  journal = {Physical Review D}, volume = {41}, pages = {1783--1791}, year = {(1990)}
}

@article{Kaiser2010,
  author = {Kaiser, David I.},
  title = {Conformal transformations with multiple scalar fields},
  journal = {Physical Review D}, volume = {81}, pages = {084044}, year = {(2010)},
  doi = {10.1103/PhysRevD.81.084044}
}

@article{FaraoniGunzigNardone1999,
  author = {Faraoni, Valerio and Gunzig, Edgard and Nardone, Pasquale},
  title = {Conformal transformations in classical gravitational theories and in cosmology},
  journal = {Fundamental Cosmology and Physics}, volume = {20}, pages = {121--175}, year = {(1999)}
}

@article{Bostan2020,
  author = {Bostan, Nilay},
  title = {Palatini Higgs and Coleman--Weinberg inflation with non-minimal coupling},
  year = {(2020)}, eprint = {2009.04406}, archivePrefix = {arXiv}, primaryClass = {gr-qc}
}

@article{BostanRoyChoudhury2024,
  author = {Bostan, Nilay and Roy Choudhury, Shouvik},
  title = {First constraints on non-minimally coupled natural and Coleman--Weinberg inflation and massive neutrino self-interactions with {Planck}+{BICEP}/{Keck}},
  journal = {Journal of Cosmology and Astroparticle Physics}, volume = {2024}, number = {07}, pages = {032}, year = {(2024)},
  eprint = {2310.01491}, archivePrefix = {arXiv}, primaryClass = {astro-ph.CO}
}

@article{LiddleParsonsBarrow1994,
  author = {Liddle, Andrew R. and Parsons, Paul and Barrow, John D.},
  title = {Formalizing the slow-roll approximation in inflation},
  journal = {Physical Review D}, volume = {50}, pages = {7222--7232}, year = {(1994)},
  doi = {10.1103/PhysRevD.50.7222}
}

@article{StewartLyth1993,
  author = {Stewart, Ewan D. and Lyth, David H.},
  title = {A more accurate analytic calculation of the spectrum of cosmological perturbations produced during inflation},
  journal = {Physics Letters B}, volume = {302}, pages = {171--175}, year = {(1993)}
}

@article{LiddleLeach2003,
  author = {Liddle, Andrew R. and Leach, Samuel M.},
  title = {How long before the end of inflation were observable perturbations produced?},
  journal = {Physical Review D}, volume = {68}, pages = {103503}, year = {(2003)}
}

@article{Planck2020,
  author = {{Planck Collaboration} and Akrami, Y. and others},
  title = {Planck 2018 results. X. Constraints on inflation},
  journal = {Astronomy \& Astrophysics}, volume = {641}, pages = {A10}, year = {(2020)},
  doi = {10.1051/0004-6361/201833887}, eprint = {1807.06211}
}

@article{BK2021,
  author = {Ade, P. A. R. and others},
  collaboration = {BICEP and Keck},
  title = {Improved constraints on primordial gravitational waves using {Planck}, {WMAP}, and {BICEP}/{Keck} observations through the 2018 observing season},
  journal = {Physical Review Letters}, volume = {127}, pages = {151301}, year = {(2021)},
  doi = {10.1103/PhysRevLett.127.151301}
}

@article{Aiola2020,
  author = {Aiola, Simone and others},
  collaboration = {ACT},
  title = {The {Atacama Cosmology Telescope}: DR4 maps and cosmological parameters},
  journal = {Journal of Cosmology and Astroparticle Physics}, volume = {2020}, number = {12}, pages = {047}, year = {(2020)},
  doi = {10.1088/1475-7516/2020/12/047}, eprint = {2007.07288}
}

@article{LewisBridle2002,
  author = {Lewis, Antony and Bridle, Sarah},
  title = {Cosmological parameters from {CMB} and other data: A {Monte Carlo} approach},
  journal = {Physical Review D}, volume = {66}, pages = {103511}, year = {(2002)},
  doi = {10.1103/PhysRevD.66.103511}, eprint = {astro-ph/0205436}
}

@book{FujiiMaeda2009,
  author = {Fujii, Yasunori and Maeda, Kei-ichi},
  title = {The Scalar-Tensor Theory of Gravitation},
  series = {Cambridge Monographs on Mathematical Physics},
  publisher = {Cambridge University Press},
  address = {Cambridge},
  year = {(2009)},
  doi = {10.1017/CBO9780511535093}
}

\end{document}